\documentclass[11pt,a4paper]{article}
\usepackage{amsmath}
\usepackage{dsfont}
\usepackage{bm}
\usepackage{esint}
\usepackage{subcaption}
\usepackage{accents}
\usepackage{mathrsfs}
\usepackage{amsbsy}

\usepackage{a4wide,graphicx,times,psfrag,wrapfig,sidecap}
\usepackage{cite}
\usepackage[colorlinks=true,linkcolor=black, citecolor=black,
urlcolor=black]{hyperref}
\numberwithin{equation}{section}
\makeatletter \let\old@startsection=\@startsection
\renewcommand{\@startsection}[6]
{\old@startsection{#1}{#2}{#3}{#4}{#5}{#6\mathversion{bold}}}
\makeatother

\def\<{\langle}
\def\>{\rangle}

\def\tr{{\rm   tr} }
\def\Im{{\rm Im}}
\def\Re{{\rm Re}}
\newcommand\encadremath[1]{\vbox{\hrule\hbox{\vrule\kern8pt
\vbox{\kern8pt \hbox{$\displaystyle #1$}\kern8pt}
\kern8pt\vrule}\hrule}} \def\enca#1{\vbox{\hrule\hbox{
\vrule\kern8pt\vbox{\kern8pt \hbox{$\displaystyle #1$} \kern8pt}
\kern8pt\vrule}\hrule}}
  \usepackage{bm}

\begin{document}

\begin{center}
{\Large\bf The Whitham Approach to the $c\to0$ limit of The Lieb-Liniger Model and Generalized Hydrodynamics}

\vspace{20mm}

Eldad Bettelheim   \\[7mm]

Racah Inst.  of Physics, \\Edmund J. Safra Campus,
Hebrew University of Jerusalem,\\ Jerusalem, Israel 91904 \\[5mm]

\end{center}

\vskip9mm

\vskip18mm

\noindent{ The Whitham approach is a well-studied method to describe  non-linear integrable systems. Although approximate in nature, its results may predict rather accurately the time evolution of such systems in many situations given initial conditions. A similarly powerful approach has recently emerged that is applicable to quantum integrable systems, namely the generalized hydrodynamics approach. This paper aims at showing that the Whitham approach is the semiclassical limit of the generalized hydrodynamics approach by connecting the two formal methods explicitly on the example of the Lieb-Liniger model on the quantum side to the non-linear Schr\"{o}dinger equation on the classical side in the $c\to0$ limit, $c$ being the interaction parameter. We show how quantum expectation values may be computed in this limit based on the connection established here which is mentioned above.         }

\thispagestyle{empty}
\vspace{1cm}
\setcounter{footnote}{0}
\begin{center}
\end{center}
\section{Introduction}

Classical integrability\cite{Das:Solitons,Faddeev:Book:Hamiltonian:Methods,Soliton:Book} is a venerable subject in mathematical physics enjoying a large number of useful exact results which have their roots in the  mathematical structure underlying such systems,  starting with the existence of a sufficient number of conserved quantities to completely determine the evolution of the dynamical system. Nevertheless, one of the most powerful methods of studying the initial value problem in as generic a setting as possible, but one that is still tractable, is an approximate method which is referred to, interchangeably, as the Whitham method\cite{Whitham:Book} or the dispersionless limit\cite{89:Kodama:Bloch,Carroll:Remarks:On:Whitham,88:Flaschka:KdV:Avging,Gurevich:Pitavsk}. 

The Whitham method relies on the observation that smooth initial conditions quickly break into sharp oscillatory features\cite{Gurevich:Pitavsk}. These oscillations may be understood as the system locally tracing out trajectories on invariant tori in an appropriate phase space\cite{88:Flaschka:KdV:Avging}, and are also related to the fact that the system is highly conservative, having an infinite number of conservation laws, such that oscillations have no chance of dissipating. The Whitham approach provides a way to predict the appearance of the oscillatory features and to track their shape  based on the idea that one may write equations for averaged quantities of the oscillations, called moduli of the oscillations, e.g., the  amplitude, average value, frequency, etc. of the oscillations.

To apply the Whitham method\cite{83:Krichever:Averaging}, one then first builds up a large inventory of oscillatory solutions to the nonlinear equations with {\it\ fixed} moduli. Namely solutions displaying steady oscillations, of which there are, as it turns out, a large class of tractable solutions, termed $n$-phase solutions,  $n\geq1$. One then applies the Whitham averaging method to write down evolution equations for these moduli to predict how the frequency, amplitude, average value, and so forth (additional moduli are needed for higher $n-$phase solutions), to be slowly modulate in an actual solution of the original nonlinear dynamical system. 

A very similar approach\cite{Casto:Alvaredo:Doyon:Yoshimura:GHD} has recently been  introduced in the realm of quantum integrable systems, namely an approximate  method has been devised to describe the evolution of a quantum integrable system  placed in some non-stationary  initial conditions, namely initial conditions that are not an eigenstate of the Hamiltonian. Such a system is described by an infinite number of conserved densities, the expectation values of which depend on space, in contrast to a stationary solution which is characterized by the property that the expectation values of the conserved densities are space-independent. To be able to study the evolution of the system one must assume that the conserved quantities smoothly (or rather, slowly) depend on the spatial coordinate. 

The slow evolution of the expectation values of the conserved densities is then predicted based on the continuity equation associated with each conservation law. Indeed, since the conserved currents may be calculated based on the knowledge of the conserved densities and some assumptions which will discussed in the next paragraph, the continuity equations may be written for each of the conserved densities in a closed form and then solved. These solutions are then sufficient to describe the evolution of the quantum system since the conserved quantities completely determine the behavior of the system just as in the classical case. 

The assumption underlying the ability to write an explicit form for the continuity equations is that locally the wavefunction may be approximate with a stationary state, namely an eigenstate of the Hamiltonian, which is characterized by space-independent conserved densities. For such a state the conserved currents may be written given the value of the conserved densities. Now for the spatially varying state, these expressions for the currents are simply taken from the spatially homogeneous state, but now one lets the conserved densities be space-dependent. The assumption is then that of local stationarity -- a generalization of local equilibrium. 

It is  now fairly easy to connect conceptually the Whitham method, applicable to classical systems, and the generalized hydrodynamics\cite{Casto:Alvaredo:Doyon:Yoshimura:GHD}  approach applicable to the quantum systems. Indeed, the steady oscillatory solutions of the classical system are analogous to space-independent stationary quantum states, and the modulation equations of the oscillatory solutions under the Whitham averaging method are just continuity equations for the conserved densities just as in the quantum case. 

In this paper we shall show this correspondence on a formal, rather than conceptual, level, by applying the Whitham method to a classical system one hand and the generalized hydrodynamics method to a quantum system on the other and then verifying that the two methods agree in the semiclassical limit. Namely, they produce the same solutions in the limit. The quantum system we shall study is the Lieb-Liniger model and the classical system will be the non-linear Schr\"{o}dinger equations, the former tending to the latter in a certain limit (the $c\to0$ limit, $c$ being the Lieb-Liniger interaction constant). 

Utilizing the connection established we shall study the behavior of expectation values in the quantum regime, especially at points where the classical equations begin to oscillate. We find, that although the quantum averaging also averages over  the rapid classical oscillations, ripples appear at the the average momentum fluctuation  at the points where oscillations break out, much in the same way as these occur in free fermions\cite{Bettelheim:Glazman}, which happens to be the $c\to\infty$ limit of the Lieb-Liniger model. Namely in the directly opposite to the semiclassical one. Nevertheless, the classical oscillations and the quantum ripples are of a very different nature. Indeed, as mentioned, the classical oscillations are completely wiped out by the quantum averaging, and what remains are  ripples that are of a different origin. 
\section{Classical and Quantum Integrability}
We start by the formalism which will be necessary in the rest of the paper. First we review classical integrability and the algebro-geometrical approach to it in the next subsection. The subsection after that will review the necessary material from quantum integrability. This review is mainly necessary in order to  establish notations that are the most convenient to see the analogy between the two approaches (the quantum and classical ones).
\subsection{Basic Tenets of the Algebro-geometrical Approach\label{SectionAlgebricGeometrical}}
The classical system that this paper shall be concerned with is the nonlinear Schr\"{o}dinger equation:
\begin{align}
\imath\hbar  \partial_t\psi=-\frac{\hbar^2}{2m}        \partial_x^2\psi+ c\bar\psi \psi^2.\label{NLS} 
\end{align}
The solution of this equation follows closely the solution of the Lieb-Liniger model through the Bethe approach. The analogy is best seen when one solves the nonlinear Schr\"{o}dinger equation by the Algebro-geometrical approach which ultimately leads also to separated variables. We thus describe this approach here, as it applies to the nonlinear Schr\"{o}dinger equation\cite{belokolos:Bobenko:Algebro:Geometrical:Integrable}.

The algebro-geometrical approach\cite{132:Krich:Intgr:AlgGeo,84:Dubrovin:Algebr:Geome,89:Kodama:Bloch,belokolos:Bobenko:Algebro:Geometrical:Integrable} allows one to construct a large number of multi-phase oscillatory solution of the nonlinear equation. A $g$-phase solution is labeled by a set of real numbers $\lambda_i$, which we order $\lambda_1<\lambda_2<\lambda_3<\dots<\lambda_{2g+2}$. These numbers serve as moduli.  

The solution starts by defining a mathematical object named the matrix valued Baker-Akhiezer function, the domain of which is a hypergeometric Riemann surface with the moduli serving as the branch points: 
\begin{align}
y^2=\prod_{i=1}^{2(g+1)}(\lambda-\lambda_i).\label{RiemannSurface}
\end{align}
It is convenient at this point to define a set $\mathcal{S}$:
\begin{align}
\mathcal{S}=[\lambda_1,\lambda_2]\cup[\lambda_3,\lambda_4]\cup\dots\cup[\lambda_{2g+1},\lambda_{2g+2}].
\end{align}
  Next define a set of cycles $a_i$ and $b_i$ such that $b_i$ encircles the cuts $[\lambda_1,\lambda_2]\cup\dots\cup[\lambda_{2i-1},\lambda_{2i}]$ on the upper sheet counterclockwise while the cycle $a_i$ goes clockwise from a point on the cut $[\lambda_{2i-1},\lambda_{2i}]$ to a point on the cut $[\lambda_{2i+1},\lambda_{2i+2}]$ and then returns on the lower sheet to the original point to close the cycle.

The Baker-Akhiezer function is then the unique function that obeys the following analytical conditions:
\begin{itemize}
\item{} $\Psi(\lambda)$ is a two by two matrix with  meromorphic valued entries  on the Riemann surface given by Eq. (\ref{RiemannSurface}) except at infinities on either sheets (these infinities are denoted by  $\infty_\pm$ where the index refers to the upper or lower sheet, respectively) where it has an essential singularity. At infinity on the upper sheet the behavior is prescribed to be:
\begin{align}
\Psi(\lambda)=\left(\mathds{1}+\sum_{k=1}^\infty\frac{\bm{\Phi}^{(k)}(x,t)}{\lambda} +O(\lambda^{-2})\right)e^{ \frac{\imath\sigma_z}{\hbar}\left(\lambda x-\frac{\lambda^2 }{m}t\right)} \left(\begin{array}{cc}0  &1\\ \alpha \lambda & 0\end{array}\right),\label{BAasymptotics}
\end{align}
where $\mathds{1}$ and $\boldsymbol{\Phi}^{(k)}$ are two by two matrices, the former matrix being of course  the identity. The behavior on the lower sheet is determined by the condition: $\Psi(\lambda_-)=\Psi(\lambda_+)\sigma_x,$ where again the index on $\lambda_\pm$ refers to whether the point $\lambda$ is placed on the upper or lower sheet, respectively.
\item{Away from infinity the Baker-Akhiezer function has poles at an immovable (time and space independent) set of points given by $D_i$}
\end{itemize}

A function that satisfies the above conditions is unique as can be shown using theorems from the theory of analytic functions on algebraic Riemann surfaces. This uniqueness allows one to prove the following relations:\begin{align}
&\left(\hbar\partial_x -\imath \sigma_z\lambda+\imath[\sigma_z,\bm{\Phi}^{(1)}]\right)\Psi=0, \quad \hbar \partial_x\bm{\Phi}^{(1)}+\imath [\sigma_z,\bm{\Phi}^{(1)}]\bm{\Phi}^{(1)}-\imath[\sigma_z,\bm{\Phi}^{(2)}]=0\nonumber \\
&\left(m\hbar \partial_t+\imath\sigma_z \lambda^2-\imath\lambda[\sigma_z,\bm{\Phi}^{(1)}]+\imath[\sigma_z,\bm{\Phi}^{(1)}]\bm{\Phi}^{(1)}-\imath[\sigma_z,\bm{\Phi}^{(2)}]\right)\Psi=0\label{RawUV}
\end{align} 
Indeed, one can show that adding the left hand side of any of the equations above to the Baker-Akhiezer function does not change its analytical properties, thus in order for uniqueness to hold these expressions on the left hand side must all be zero.  The equations above may be written as follows:
\begin{align}
(\hbar\partial_x - V)\Psi =0 , \quad (m\hbar\partial_t -U)\Psi=0,\label{LinearProblems}
\end{align}
where: 
\begin{align}
V=\left(\begin{array}{cc}
\imath \lambda & \imath\sqrt{c} \bar\psi \\
-\imath\sqrt{c}\psi & -\imath \lambda
\end{array} \right),\quad U=\left(\begin{array}{cc}
-\imath\lambda^2-\frac{\imath c\bar \psi \psi}{2}  &- \imath\lambda  \sqrt{c} \bar\psi-\frac{\sqrt{c}\hbar}{2}\partial_x\bar \psi \\
 \imath\lambda  \sqrt{c} \psi-\frac{\sqrt{c}\hbar}{2}\partial_x\psi & \imath\lambda^2+\frac{\imath c\bar \psi \psi}{2}
\end{array} \right)\label{UV}
\end{align}
and  $\Phi^{(1)}_{12}=-\frac{\sqrt{c} \bar \psi}{2}$ and $\Phi^{(1)}_{21}=-\frac{\sqrt{c}  \psi}{2}.$ The derivation of Eqs. (\ref{UV}) from Eqs. (\ref{RawUV}) is straightforward, the only care that need to be taken is in separating the diagonal and off-diagonal components of $[\sigma_z,\bm{\Phi}^{(1)}]\bm{\Phi}^{(1)}$, the former being expressible through the off-diagonal components of $\bm{\Phi}^{(1)}$ and the latter begin determined from the second equation in Eqs. (\ref{RawUV}). 

For the two equations in (\ref{LinearProblems}) to be compatible, the zero curvature condition must hold, $[m\hbar \partial_t-U,\hbar\partial_x-V]=0,$ which, in turn, requires the Non-linear Schr\"odinger equation, Eq. (\ref{NLS}), as can easily be confirmed by direct computation. 

Thus starting with the moduli we have ended up with a solution of the nonlinear Schr\"{o}dinger equation, provided that an explicit expression for the Baker-Akhiezer function can be written such that the $12$ element of $\boldsymbol{\Phi}^{(1)}$, which is equal to the nonlinear Schr\"{o}dinger field $-\frac{\sqrt{c}\bar\psi}{2}$, may be extracted. The well-known explicit expression for the Baker-Akhiezer function based on its analytic properties alone will be presented below, but we shall first be interested in obtaining  the conserved densities and the conserved currents from the algebro-geometrical approach, as these will serve as the link between the classical and quantum system. Namely, we shall identify the semiclassical limit of a certain quantum eigenstate as a classical solution having the same values of the conserved densities as the expectation values of these quantities in  the corresponding eigenstate.

To the aim of extracting the conserved densities from the Baker-Akhiezer function we define, as is customary, the propagator of the evolution $\boldsymbol{T}(x,t;x',t')$. The propagator is defined as a matrix, the columns of which  solve (\ref{LinearProblems}) as a function of $x'$ and $t'$, while having  initial conditions $\bm{T}(x,t;x,t)=\mathds{1}.$ The propagator is also a function of the spectral parameter $\lambda$. Since the propagator  obeys the same matrix differential equations as the Baker-Akhiezer function, Eq. (\ref{LinearProblems}), its columns must be linear combinations of the columns of the Baker-Akhiezer function. Namely,
\begin{align}
\bm{T}(x,t;x',t')= \Psi(x',t') \left( \begin{array}{cc} 
\alpha & \gamma  \\
\beta & \delta 
 \end{array}\right)
\end{align} where $\alpha,$ $\beta,$ $\gamma$ and $\delta$ are functions of the spectral parameter, $\lambda,$ and of $x$ and $t$ but not of $x'$ and $t'$.  The parameters $\alpha,$ $\beta,$ $\gamma$ and $\delta$  will be found  below in the appendix based on the explicit form of the Baker-Akhiezer function. 

The equal time propagator and with spatial arguments differing by the basic spatial period, $\boldsymbol{T}(x,t;x+L,t)$ is called the monodromy matrix. 

If an oscillatory solution has a spatial period $L$ and a temporal period $T$ then one may easily derive:
\begin{align}
\partial_t \tr(\bm{T}(x,t;x+L,t)=0, \quad \partial_x\tr(\boldsymbol{T}(x,t;x,t+T)=0.\label{trTconserves}
\end{align}
which will be shown below to be relations allowing to establish the trace of the propagator as a generating function for conserved densities and currents. 

To obtain relations (\ref{trTconserves}) one writes 
\begin{align}
\boldsymbol{T}=\mathcal{P}e^{\frac{1}{\hbar}\int_{x,t}^{x',t'} (Vdx+U\frac{dt}{m}) },
\end{align}
where $\mathcal{P}$ denotes path ordering of the matrix exponent and the right hand side is well defined due to the zero curvature condition. This representation allows one to write:
\begin{align}
&\partial_t\boldsymbol{T}(x,t;x+L,t)=U(x,t)\boldsymbol{T}(x,t;x+L,t)-\boldsymbol{T}(x,t;x+L,t)U(x+L,t),\label{XMonodromy}\\&\partial_x \boldsymbol{T}(x,t;x+L,t)=V(x,t)\boldsymbol{T}(x,t;x,t+T)-\boldsymbol{T}(x,t;x,t+T)V(x,t+T)\label{TMonodromy}.
\end{align}
Letting $x'=x+L,$ $t'=t$ for $L$ a period and taking the trace gives the first of the relations in (\ref{trTconserves}) while taking taking $x'=x$, $t'=T+L$ for $T$\ a period and taking the trace gives the second relation in (\ref{trTconserves}).

 As mentioned above, the  explicit construction of the Baker-Akhiezer function based on its analytic properties on the Riemann surface alone, allows one also to obtain explicit expressions for the matrix elements of $\boldsymbol{T}$, which in turn allows one to compute the traces in (\ref{trTconserves}). The final result, to be confirmed in the sequel, is given by: 
\begin{align}
&\tr(\bm{T}(x,t;x+L,t)=2\cos\left(  \frac{L}{\hbar } \Omega^{(c)}_0(\lambda)\right),\label{trTisCos}\\
&\tr(\bm{T}(x,t;x,t+T)=2\cos\left(  \frac{2T}{m\hbar} \Omega^{(c)}_1(\lambda)\right),\label{trTisCos1}
\end{align}
where $\Omega^{(c)}_0$ is the integral over the differential $d\Omega^{(c)}_i,$ which are defined as the unique differential which  have a poles of order $i+2$  at infinity at the upper and lower sheets of the Riemann surface with residue equal to $\pm1$ respectively and which have vanishing $a$ cycles. These conditions read formally as: 
\begin{align}
d\Omega_i\sim\pm d\lambda \lambda^{i}\mbox{\,\,   for\,\, }\lambda\to\infty_\pm,\mbox{  and } \quad \oint_{a_j} \Omega_i=0.\label{Omegaidef}
\end{align} The superscript $(c)$ in $\Omega_i^{(c)}$ denotes that this is a classical object, a distinction that will be important to make since we shall encounter the same object in the quantum case as well, and it will be important to differentiate the two in the notations. 

One may find an explicit expression for  $d\Omega^{(c)}_i.$ It has the form:
\begin{align}
&d\Omega^{(c)}_0=\frac{\lambda^{g+1}-\frac{\sum_{i}{\lambda_i}}{2}\lambda^g+\sum_{j=0}^{g-1}c^{(0)}_j\lambda_j}{\sqrt{\prod_i (\lambda-\lambda_i)}}\label{dOmega0Explicit}d\lambda,\\
&d\Omega^{(c)}_1=\frac{\lambda^{g+2}-\frac{\sum_{i}{\lambda_i}}{2}\lambda^{g+1}-\frac{1}{4}\sum_{i}\lambda_i\left(\lambda_i-\frac{\sum_j\lambda_j}{2}\right) +\sum_{j=0}^{g-1}c^{(1)}_j\lambda_j}{\sqrt{\prod_i (\lambda-\lambda_i)}}d\lambda ,\label{dOmega1Explicit}
\end{align}
where the $c^{(0,1)}_j$ are determined by the condition that the $a-$cycles of $d\Omega^{(c)}_{0,1}$vanish, respectively. For example for $g=1$ these are expressible as elliptic integrals.

Due to the conservation, Eq. (\ref{trTconserves}), of properly defined traces of the propagator for any value of the spectral parameter, $\lambda$,  we may define conserved charges and currents based on these traces. These are defined as follows:
\begin{align}
&\frac{-\imath \hbar}{L}\log\tr\,\boldsymbol{T}(x,t;x+L,t)\overset{\lambda\to-\imath\infty}{=} \Omega^{(c)}_0 (\lambda)=\lambda+\frac{1}{L}\sum_{k=1}^\infty I_k \lambda^{-k-1} ,\label{IkThroughLogT}\\
&\frac{-\imath \hbar }{T}\log\tr\,\boldsymbol{T}(x,t;x,t+T)\overset{\lambda\to\sqrt{-\imath}\infty}{=} \frac{2}{m}\Omega^{(c)}_1 (\lambda)=\frac{\lambda^2}{m}-\frac{1}{L}   \sum_{k=1}^\infty J_k \lambda^{-k-1} ,\label{JkThroughLogT}
\end{align} 
where equalities are up to factors exponentially small in $\lambda$ in the limit and  $I_k,$ $J_k$ are conserved charges  currents, respectively,  obeying the following equations as a consequence of their definition and of Eqs. (\ref{trTconserves}): 
\begin{align}
\partial_t I_k=0, \quad \partial_x J_k=0,
\end{align}
for any $k$. These conserved quantities may be written then as:
\begin{align}
&I_k =L \oint_{\infty}\lambda^{k} \Omega^{(c)}_0\frac{d\lambda}{2\pi\imath}=L\oint_{\mathds{R}} \lambda^{k} \Omega^{(c)}_0\frac{d\lambda}{2\pi\imath} \label{IkClassical} \\
& J_k =- \frac{2L}{m}\oint_{\infty}\lambda^{k} \Omega^{(c)}_1\frac{d\lambda}{2\pi\imath}=-\frac{2L}{m}\oint_{\mathds{R}} \lambda^{k} \Omega^{(c)}_1\frac{d\lambda}{2\pi\imath},\label{JkClassical} 
\end{align}
where the integrals on the right hand sides of both equations are to be taken around the real axis or, equivalently, only  around the set $\mathcal{S}$, where $\Omega^{(c)}_i$ have jump discontinuities. These quantities can then be computed by knowledge of $\lambda_i$ alone, namely they depend only on the moduli of the oscillatory solution, or equivalently on the algebraic Riemann surface. 

Note that $I_k$ must be thought of as related to an underlying conserved density $\rho_k$, as it is averaged over a period of the oscillation and  $J_k$ is a conserved current density similarly averaged:
\begin{align}
I_k = \int_{x}^{x+L} \rho_k, \quad J_k=\int_{x}^{x+L} j_k.\label{IJthroughpj}
\end{align} 
The densities and current densities then obey a continuity equation
\begin{align}
\partial_t \rho_k=\partial_x j_k,\label{continuitydens}
\end{align}
which should seem  plausible at this point, but we shall  not show this here, referring the reader to the appropriate textbooks and manuscripts instead\cite{belokolos:Bobenko:Algebro:Geometrical:Integrable,88:Flaschka:KdV:Avging}. We should mention that a crucial part of the identification of the conserved quantities, $I_k,$  as giving the dynamical system an integrable structure is that they must be shown to be in involution. To that end one may show that the monodromy matrix for different values of the spectral parameter are in involution ( commute under the Poisson bracket). We also refer the reader to the references above for the proof of this fact. Explicitly what may be proven is the following relation
\begin{align}
\{\mathcal{T}(\lambda),\mathcal{T}(\mu)\}=0 \label{classicalallcommutes}
\end{align}    
where
\begin{align}
\mathcal{T}=\tr(\hat{\boldsymbol T}(x,t;x+L,t)).
\end{align}

\subsection{Basic Tenets of the Quantum Inverse Scattering Method}
We wish to associated, in the semiclassical limit, with each eigenstate of a quantum Hamiltonian (the semiclassical limit of which is the nonlinear Schr\"{o}dinger equation's Hamiltonian) a classical oscillatory solution. The quantum system in question is quantum integrable, which means that there are an infinite number of operators commuting with the Hamiltonian such that with each eigenstate   we may associated an infinite number of eigenvalues $\{I^{(Q)}_k\}_{k=1}^\infty$, namely the eigenvalues  with respect to these conserved operators. The classical solution has, on the other hand, an infinite number of conserved charges defined in (\ref{IkThroughLogT}) which characterize it, denote this set also by $\{I^{(c)}_k\}_{k=1}^\infty. $ The classical solution and the quantum eigenstate are associated with one another if the two sets above contain  the same values for corresponding conserved quantities  $\{I^{(Q)}_k\}_{k=1}^\infty=\{I^{(c)}_k\}_{k=1}^\infty$. We thus discuss now how the conserved charges appear in the quantum settings within the quantum inverse scattering method, as this method has the most direct relation to the algebro-geometrical approach we have used in the pervious section to discuss the classical case.

One may arrive at the quantum non-linear Schr\"{o}dinger equation as follows (here we follow closely the book of Ref.\cite{Korepin:Bogoliubov:Izergin:Quantum:Inverse:Scattergin}). Rather than starting with a Hamiltonian, one first defines a quantum matrix   $\hat{V}$:
\begin{align}
 \hat{V}(x,t)=\left(\begin{array}{cc}
\imath \lambda & -\imath\sqrt{c} \hat\psi^\dagger \\
\imath\sqrt{c}\hat\psi & -\imath \lambda
\end{array} \right),
\end{align}
where $\hat\psi$ is a canonical bosonic field operator. Then one defines a spatial propagator as follows:
\begin{align}
&\partial_x\hat{\boldsymbol{T}}(x,x')=\hat{V}(x,t)\hat{\boldsymbol{T}}(x,x'), \quad \hat{\boldsymbol{T}}(x,x)=\mathds{1},
\end{align}
the solution of which is 
\begin{align}
\hat{\boldsymbol T}=\mathcal{P}e^{\frac{1}{\hbar}\int_x^{x'} \hat{V}(y)dy}.\label{QTisExp}
\end{align}
This is followed by a definition of an operator which represents the trace of the monodromy matrix :
\begin{align}
\hat{\mathcal{T}}(\lambda)=\tr (\hat{\boldsymbol T}(x,x+L,t)).
\end{align}
One may prove that for any $\lambda$ and $\mu$ the following, a quantum analogue of Eq. (\ref{classicalallcommutes}), holds:
\begin{align}
[\hat{\mathcal{T}}(\lambda),\hat{\mathcal{T}}(\mu)]=0,\label{quantumallcommutes}
\end{align}
 based only on the definition of $\hat{V}$ and the bosonic commutation relations of the field $\hat{\psi}$.

From Eq. (\ref{quantumallcommutes}) one may define an infinite set of mutually commuting conserved charges:
\begin{align} 
-\frac{\imath \hbar}{L}\log(\hat{\mathcal{T}}(\lambda))\overset{\lambda\to\imath\infty}{=}\lambda  -\frac{\imath \hbar}{L}+\frac{1}{L}\sum_{k=1}^\infty \hat{I}_k \lambda^{-k-1},\label{QlogTisI}
\end{align}
to be compared to Eq. (\ref{IkThroughLogT}).  To compute $\hat{I}_k$ , according to this equation one must only exponentiate the integral of $\hat{V}$ in a path ordered fashion according Eq. (\ref{QTisExp}) and expand the logarithm of the result. The outcome of this calculation for the first three conserved charges is then:
\begin{align}
\hat{I}_0=\frac{\imath \hbar c}{2}\int \hat{\rho},\quad \hat{I}_1=\frac{\imath\hbar c}{4}\int \hat{\mathcal P} +O(c^2), \quad \hat{I}_2=\frac{\imath\hbar cm}{4}\int \hat{\mathcal{H}}+O(c^2) ,
\end{align}
where $\hat \rho$ , $\hat{\mathcal{P}}$ and $\hat{\mathcal{H}}$ are the particle, momentum and Hamiltonian density, respecitvely:
\begin{align}
\hat{\rho}=\hat\psi^\dagger \hat\psi, \quad \hat{\mathcal P}=-\imath\hbar\hat\psi^\dagger  \partial_x\hat\psi, \quad \quad \hat{\mathcal H}=\frac{\hbar^2}{2m}\partial_x \hat\psi^\dagger \partial_x \hat\psi+\frac{c}{2} \hat\psi^\dagger\hat \psi^\dagger\hat\psi\hat\psi,
\end{align}
the latter being the Hamiltonian for the quantum nonlinear Schr\"{o}dinger equation.

One may now simultaneously diagonalize the Hamiltonian with all other conserved quantities, or in other words find a common eigenvector for $\hat{\mathcal{T}}(\lambda)$ simultaneously for all values of the spectral parameter. The eigenvalue is denoted by $t(\lambda).$ The algebraic Bethe ansatz solves the problem of finding the eigenvector and eigenvalue in question. 

We shall not go over in any detail a description of the algebraic Bethe ansatz, but we shall only remind the basic ideas behind it, since the semiclassical analogue of the objects that appear there will bear some importance for us in the sequel. 

To proceed with the diagonalization of the trace of the monodromy matrix, it is actually useful to consider all four elements of the matrix rather than just the trace. These are denoted as follows:
\begin{align}
\hat{\boldsymbol{T}}(\lambda)=\left(\begin{array}{cc} \hat{A}(\lambda)& \hat{B}(\lambda)\\
\hat{C}(\lambda) &\hat{D}(\lambda) \end{array}\right)
\end{align} 
It is possible to write down the full algebra of the operator $\hat{A},$ $\hat{B}$, $\hat{C}$ and $\hat{D}$ giving the commutation relations between them at different values of the spectral parameters. The list of these commutation relations is quite long, so we shall not repeat it here, noting only the Abelian nature of the operators $\hat{B}$:
\begin{align}
[\hat B(\lambda),\hat B(\lambda')]=0.
\end{align}
Crucially, it turns out that the eigenvectors of $\hat{\mathcal{T}}=\tr(\hat{\boldsymbol{T}}), $ denoted by $|\{\theta_ i\}_{i=1}^N\>$ may be written through the operators $\hat{B}$ as follows:
\begin{align}
|\{\theta_i\}_{i=1}^N\>=\hat{B}(\theta_1)\hat{B}(\theta_2)\dots\hat{B}(\theta N)|0\>, 
\end{align}
where the set $\{\theta_i\}_{i=1}^N$ is called the set of Bethe roots and has to be chosen appropriately in order for the vector $|\{\theta_i\}_{i=1}^N\>$  to be an eigenstate of $\hat{\mathcal{T}}$.  Namely, the roots have to obey the Bethe equations:
\begin{align}
e^{-\frac{\imath2\theta_i L}{\hbar}}\prod_{j= 1}^N\frac{\theta_i-\theta_j+\imath c/2}{\theta_i-\theta_j-\imath c/2}=-1.\label{BetheEqs}
\end{align}
Given such a choice of roots, namely given a solution of the Bethe equations, the eigenvalue of $\hat{\mathcal{T}}(\lambda)$, denoted by $t(\lambda)$ is given by:
\begin{align}
t(\lambda)=e^{-\frac{\imath \lambda L}{\hbar}}\prod_{j=1}^N \frac{\lambda-\theta_j+\imath c/2}{\lambda -\theta_j}+e^{\frac{ \imath \lambda L}{\hbar} }\prod_{j=1}^N \frac{\lambda-\theta_j-\imath c/2}{\lambda-\theta_j},\label{tFromBethe} 
\end{align}
The eigenvalue of the conserved charges $\hat{I}_k$ denoted by $I_k$ is then computed by invoking  Eq. (\ref{QlogTisI}) and Eq. (\ref{tFromBethe}) which yields 
\begin{align}
 I_k=\frac{\hbar  c}{2}\sum_j \theta_j^{k}.\label{QIasFofRoots}
\end{align} 

We note that the Bethe equations ensure that, as a function of $\lambda$, $t(\lambda)$ does not have poles on the Bethe roots $\lambda_i$, even though glancing at (\ref{tFromBethe}), there is a pole singularity at each of the roots. Indeed, the Bethe equations are simply the equations that ensure that the residue vanishes at these poles. 

Comparing Eq. (\ref{trTisCos}) and Eq. (\ref{tFromBethe})\ allows one to obtain an equation for an object we shall denote as,  $\Omega^{(Q)}_0,$ which is the quantum analogue of $\Omega_0^{(c)}$, and is defined in analogy to Eq. (\ref{trTisCos}) as follows: 
\begin{align}
&t(\lambda)|\{\theta_i\}_{i=1}^N\>\equiv \tr(\hat {\bm{T}})|\{\theta_i\}_{i=1}^N\>=2\cos\left(  \frac{L}{\hbar } \Omega^{(Q)}_0(\lambda)\right)|\{\theta_i\}_{i=1}^N\>,
\end{align}
and which may be written as  function of the Bethe roots (Eq. (\ref{Omega0Throughroots}) below) that will be useful in the following. For that purpose we define  $\Lambda(\lambda)$ (which will also prove useful in and on itself) as follows: 
\begin{align}
\Lambda(\lambda)=e^{\frac{\imath \lambda L}{\hbar}}\prod_{j=1}^N \frac{\lambda-\theta_j-\imath c/2}{\lambda-\theta_j}.\label{LambdaDef}
\end{align}
Then the eigenvalue of the transfer matrix, $t(\lambda)$ can be written through $\Lambda$ from Eq. (\ref{tFromBethe}) as follows:
\begin{align}
t(\lambda)=\Lambda(\lambda)+e^{-\frac{cL}{2\hbar}   }\Lambda^{-1}(\lambda+\imath c/2).\label{tThroughLambda}
\end{align} 
As such, and comparing with (\ref{trTisCos}) one obtains that in the semiclassical limit $c\to0$ one has:
\begin{align}
\Lambda\to e^{\frac{\imath L \Omega^{(Q)}_0}{\hbar}}\label{LambdaLimit}
\end{align}
Thus $-\frac{\imath\hbar }{L}\log(\Lambda)\to\Omega^{(Q)}_0$. Taking the logarithm of (\ref{LambdaDef}) then yields:
\begin{align}
\Omega^{(Q)}_0=\lambda -\frac{c\hbar}{2L} \sum \frac{1}{\lambda -\theta _j}.\label{Omega0Throughroots}
\end{align} 

In the next section we shall show that the semiclassical expression, Eq. (\ref{Omega0Throughroots}), for $\Omega^{(Q)}_0$ with $\theta_i$ obeying the Bethe equations, Eq. (\ref{BetheEqs}), agrees with the classical expression for $d\Omega^{(Q)}_0$ as a meromorphic differential of a specific type, Eq. (\ref{dOmega0Explicit}).

\section{Convergence of the Bethe Spectrum to the Algebro-Geometric Spectrum}
We have characterized oscillatory solutions of the classical nonlinear Schr\"{o}dinger equation using a spectral surface. Namely, a Riemann surface with branch points $\lambda_i$. The quantum analogue of the oscillatory solutions are eigenstates of the quantum Hamiltonian described by Bethe roots, $\theta_i$, satisfying the Bethe equations. We now show how, in the $c\to0$ limit we may identify a spectral surface with a particular solution of the Bethe equations. Our starting point for the identification is that we require the eigenvalue of the quantum conserved charges on the eigenstate to be equal to the conserved charges of the classical state, averaged over a period of the oscillations.

Having established as the basis of the identification the conserved charges, we compare the quantum Eq. (\ref{QIasFofRoots}) to the classical Eq. (\ref{IkClassical}). If we define the density of Bethe roots as $\sigma_p\equiv \sum \frac{1}{L} \delta(\lambda-\lambda_i)$ then Eq. (\ref{QIasFofRoots})   can be written as:
\begin{align}
I^{(Q)}_k=\frac{\hbar c}{2}  \int \sigma_p\lambda^k.
\end{align}
similarly defining $\sigma^{(c)}$ as the jump discontinuity of $d\Omega^{(c)}_0$ over the real axis, $2\pi\imath \sigma^{(c)}=\Omega^{(c)}_0(x+\imath0^+)-\Omega^{(c)}_0(x+\imath0^-)$ then Eq. (\ref{IkClassical}) can be written as:
\begin{align}
I_k^{(c)}= \int  \sigma^{(c)} \lambda^k.
\end{align}
Comparing the two equations we see that we must have that the jump discontinuity of $\frac{\hbar c}{4\pi\imath}\Omega^{(c)}_0$ be equal to the density of Bethe roots, $\hbar c\sigma^{(c)}=\sigma_p,$ in order for the oscillatory solution to be identified with the solution of the Bethe equation.

The function $\sigma^{(c)}$, namely, the jump discontinuity of $\frac{1}{2\pi\imath}\Omega^{(c)}_0,$ is a continuous function supported on the set $[\lambda_1,\lambda_2]\cup[\lambda_3,\lambda_4]\cup\dots\cup[\lambda_{2g+1},\lambda_{2g+2}],$ as can be seen from Eq. (\ref{dOmega0Explicit}), such that for finite $g$, the density of the corresponding Bethe roots must be  $\sigma_p$ must also be a smooth function with finite support, containing characteristic gaps. Such a situation can only occur if the number of roots, $N,$\ goes to infinity, which requires to solve the Bethe equations in the thermodynamic limit, namely one must employ the thermodynamic Bethe ansatz, which concerns itself with the $N\to\infty$ limit.
We describe this method briefly below.

In the thermodynamic limit, the roots may be described by their density on the real axis $\sigma$. To find an equation for $\sigma$ in the thermodynamic limit one first takes the logarithm of the Bethe equations, Eq. (\ref{BetheEqs}):
\begin{align}
\pi(2j_i+1)=\frac{2\theta_i L}{\hbar}+\imath\sum_j  \log\frac{\theta_i-\theta_j+\imath c/2}{\theta_i-\theta_j-\imath c/2} \label{logBetheEqs}  
\end{align}
Subtracting the equation for $i+1$ from the equation for $i,$ while denoting $\sigma_p (\theta_i)=\frac{1}{L} \frac{1}{\theta_{i+1}-\theta_i},$  $\sigma_h(\theta_i)=\frac{1}{L}\frac{j_{i+1}-j_i-1}{\theta_{i+1}-\theta_{i}}$ and $\sigma_s(\theta_i)=\sigma_p(\theta_i)+\sigma_h(\theta_i)=\frac{1}{L}\frac{j_{i+1}-j_i}{\theta_{i+1}-\theta_{i}}$ one obtains:
\begin{align}
2\pi  \frac{\sigma_s(\theta_i)}{\sigma_p(\theta_i)}=\frac{2}{\hbar\sigma_p(\theta_i)}+\frac{c }{L\sigma_p(\theta_i)}\sum_j \frac{1}{(\theta_i-\theta_j )^2+ c^2/4}  ,  
\end{align}
turning the sum into  an integral and rearranging gives:
\begin{align}
\sigma_s(\lambda)=\sigma_p(\lambda)+\sigma_h(\lambda)=\frac{1}{\pi\hbar}+\int \frac{c}{(\lambda-\lambda')^2+ c^2/4}  \sigma_p(\lambda')\frac{d\lambda'}{2\pi} \label{sigmas} 
\end{align} 
defining $n=\frac{\sigma_p}{\sigma_s}$ gives the dressing equation:
\begin{align}
\sigma_s (\lambda)= \frac{1}{\pi\hbar }+\int \frac{c}{(\lambda-\lambda')^2+ c^2/4}  \sigma_s(\lambda')n(\lambda')\frac{d\lambda'}{2\pi}\label{SigmasDressing}
\end{align} 
which is a particular case of the dressing equation:\begin{align}
f^{\rm dr}(\lambda) = f(\lambda) +\int\frac{d\alpha}{2\pi}\frac{c}{(\lambda-\alpha)^2+c^2/4}  n(\alpha)f^{\rm dr}(\alpha)\label{dressing}
\end{align}
for $f(\lambda)=\frac{1}{\hbar \pi}$. 

Consider the dressing equation for $f(\lambda)=n(\lambda), $ where $n$ denotes the characteristic function of the union, $\mathcal{S}$, of $g+1$ segments $\mathcal{S}={\cup_{i=0}^{g}[\lambda_{2i+1},\lambda_{2i+2}]}$, namely $n=\chi_{\mathcal{S}}$. We may solve the dressing equation by decomposing $n f^{\rm dr}$ into $f^{\rm dr}_+ $ and $f^{\rm dr}_-$ as follows \begin{align}{    f^{\rm dr}_++ f^{\rm dr}_-}{} =n f^{\rm dr},\label{Jumpftilde}\end{align} where $ f^{\rm dr}_\pm$ may be analytically continued to the upper and lower half planes and behaving as $ f^{\rm dr}\pm(\lambda)=O(\frac{1}{\lambda}  )$ as $\lambda\to\infty$. With these definitions the dressing equation, Eq. (\ref{dressing}), may be represented as follows  for $\lambda \in \mathcal{S}$:
\begin{align}
&  f^{\rm dr}_+(\lambda)+ f^{\rm dr}_-(\lambda)= f(\lambda) +\int\frac{d\alpha}{2\pi}\frac{c}{(\lambda-\alpha)^2+c^2/4}  n f^{\rm dr}(\alpha)=\nonumber\\&=f(\lambda)+ f_+^{\rm dr}(\lambda+\imath c/2)+ f_-^{\rm dr}(\lambda-\imath c/2).\label{BetheInfplfmn} 
\end{align}     
Using (\ref{Jumpftilde}), one obtains: \begin{align}f_+'^{\rm dr}- f_-'^{\rm dr}=\frac{2\imath}{c} f,\label{fdrisf}\end{align}
which is valid for $\lambda\in\mathcal{S}$.

For $f=h_0\equiv \frac{1}{\pi\hbar}$, the dressing equation, Eq.(\ref{dressing}) is just the Bethe equations in the thermodynamic limit, Eq. (\ref{SigmasDressing}), and $h_0^{\rm dr}$ is just the Bethe density of states, $\sigma_s$, while $nh_0^{dr}$ is the density of Bethe roots, $\sigma_p$. The function $\Omega^{(Q)}_0$,  behaves at infinity as $\lambda$ and has jump discontinuity over the real axis given by $\pi \hbar c\imath \sigma_p$, as dictated by (\ref{Omega0Throughroots}), thus it may be written through $h_{0+}^{\rm dr}$ and $h_{0-}^{\rm dr}$\  (which are $f_\pm^{\rm dr}$ for $f=h_0=\frac{1}{\pi}$) as follows:
\begin{align}
\Omega^{(Q)}_0= \lambda+\pi c\hbar\imath  \left\{\begin{array}{lr} +h^{\rm dr}_{0+} & \Im(z)>0 \\-h^{\rm dr}_{0-} & \Im(z)<0\end{array}\right. \label{Omega0FromDressing}
\end{align}  
Thus by solving Eq. (\ref{fdrisf}) for $f=h_0$ we may compare $\Omega^{(Q)}_0$ that we obtain from the Bethe ansatz, namely Eq. (\ref{Omega0FromDressing}), and the classical expression, Eq. (\ref{dOmega0Explicit}). Showing that the classical and quantum expressions agree would mean that $\sigma_p$ is equal to the jump discontinuity of $\frac{\hbar c}{4\pi\imath} \Omega^{(Q)}_0$, which we also called $\frac{\hbar c}{2}\sigma^{(c)}$ above. We thus proceed to solve the dressing equation in order to be able to make the comparison. 

Rather than solving Eq. (\ref{fdrisf}) only for $f=h_0$, we  solve it  more generally for any $h_i=\frac{\lambda^i}{\pi\hbar\ }$  as this will also allow us to compare other quantities between the quantum and classical results. We first  generalize Eq. (\ref{Omega0FromDressing})\ and define for any $i$ the Bethe-ansatz $\Omega^{(Q)}_i$ as follows:
\begin{align}
\Omega^{(Q)}_i= \frac{\lambda^{i+1}}{i+1}+\pi c\hbar\imath  \left\{\begin{array}{lr} +h^{\rm dr}_{i+} & \Im(z)>0 \\-h^{\rm dr}_{i-} & \Im(z)<0\end{array}\right. \label{OmegaiFromDressing}
\end{align}

Consider then  $f(\lambda)=h_i(\lambda)=\frac{\lambda^i}{\pi\hbar\ }$ from. One may offer a solution to (\ref{fdrisf}) as follows:  
\begin{align}
  \pm h'^{\rm dr}_{i\pm}(\lambda)d\lambda=\frac{\imath\lambda^{i}}{\pi \hbar c} d\lambda-\imath \frac{\lambda^{i+1+g}+a^{(i)}_1 \lambda^{i+g}+a^{(i)}_2 \lambda^{i+g-1}+\dots+a_{i+g+1}^{(i)}}{\pi\hbar  c\sqrt{\prod_{j=1}^{2g+2}(\lambda-\lambda_j\pm\imath0^+ )}}d\lambda \label{htildeiexplicit}
\end{align}
where $a_{j}^{(i)}$ for $j=1,\dots,i+1$ are fixed by the requirement that $h'_{i,\pm} \sim \frac{c_i}{\lambda^2}$ for some $c_i$  as $\lambda \to \infty$, which represents $i$ conditions, while $a^{(i+2)}_{j}$ for $j=i+2,\dots,i+g+1$ (assuming $g>0$) are to be determined from an additional set of  $g$ conditions to be  identified shortly. We note that to see that  Eq. (\ref{htildeiexplicit}) indeed solves (\ref{fdrisf}) is easy as $h'^{\rm dr}_{i+}-h'^{\rm dr}_{i-}$ is easily seen to be equal to $\frac{\lambda^i}{\pi\hbar c}$ for $\lambda \in \mathcal{S}$. Furthermore this solution is consistent with Eq. (\ref{Jumpftilde}), as $h'^{\rm dr}_{i+}+h'^{\rm dr}_{i-}$ is $0$ for $\lambda\ni\mathcal{S}$, as the right hand side of (\ref{Jumpftilde}) would suggest.

Combining Eq. (\ref{htildeiexplicit})\ and Eq. (\ref{OmegaiFromDressing})\ gives:
\begin{align}
d\Omega^{(Q)}_i=\frac{\lambda^{i+1+g}+a^{(i)}_1 \lambda^{i+g}+a^{(i)}_2 \lambda^{i+g-1}+\dots+a_{i+g+1}^{(i)}}{\pi\hbar  c\sqrt{\prod_{j=1}^{2g+2}(\lambda-\lambda_j\pm\imath0^+ )}}d\lambda,\label{dOmega0ExplBethe} 
\end{align} 
which agrees with Eq. (\ref{dOmega0Explicit})\ and (\ref{dOmega1Explicit}) if the constants $a^{(i)}_j$ can be shown to agree. Indeed, these are determined by the requirement that $d\Omega^{(Q)}_i \sim \frac{d\lambda}{\lambda^2}$ as $\lambda$ tends to $\infty_\pm$ (an assumption already made above), and by the reality condition, to be discussed below, that the $a$ cycles of $d\Omega^{(Q)}_i$ must vanish. 

We now turn to the last remaining task of this section, that is, we show that the $a$-cycles of $d\Omega_i$ vanish, $\oint_{a_j} d\Omega^{(Q)}_i=0$.
In order to accomplish this, we write down an expression for $h_i^{\rm dr}$, reading\begin{align}
h_i^{\rm dr}=\left\{\begin{array}{lr}-\frac{\imath2}{\pi\hbar  c}\int_{\lambda_1}\frac{\lambda^{i+1+g}+a^{(i)}_1 \lambda^{i+g}+a^{(i)}_2 \lambda^{i+g-1}+\dots+a_{i+g+1}^{(i)}}{\sqrt{\prod_{j=1}^{g+1}(\lambda-\lambda_j +\imath 0^+)}} d\lambda& \lambda \in \mathcal{S} \\
  \frac{\lambda^{i+1+g}+a^{(i)}_1 \lambda^{i+g}+a^{(i)}_2 \lambda^{i+g-1}+\dots+a_{i+g+1}^{(i)}}{\pi\hbar c\sqrt{\prod_{j=1}^{g+1}(\lambda-\lambda_j)}} &\lambda \not\in \mathcal{S} \end{array}\right. ,\label{hisolution}
\end{align}  
which is derived as follows. For $\lambda\in\mathcal{S}$, we have $nh_i^{\rm dr}=h^{\rm dr}_{i+}+h^{\rm dr}_{i-}$, and thus the expressions for $h^{\rm dr}_{i\pm}$ must be added up from   Eq. (\ref{htildeiexplicit})\ and integrated to obtain the expression valid for $\lambda \in \mathcal{S}$. This yields the first line in Eq. (\ref{hisolution}). For $\lambda\not\in\mathcal{S},$ we may return to (\ref{dressing}), noting that the left hand of which is the function we wish to determine, while the right hand side is written through $h_i^{\rm dr}$ at the points $\lambda\in\mathcal{S}$ only, which is given in the first line of Eq. (\ref{hisolution}). In fact, for $\lambda\not\in\mathcal{S}$ the right hand side of Eq. (\ref{dressing}) reads $h_i+nh_i^{\rm  dr}+\imath c(  h'^{\rm dr}_{i+}- h'^{\rm dr}_{i-})$, where $h'^{\rm dr}_{i\pm}$  is to be analytically continued from Eq. (\ref{htildeiexplicit}). Now since $nh_i^{\rm dr}=0$, it remains to compute $h_i+\imath c(  h'^{\rm dr}_{i+}- h'^{\rm dr}_{i-})$ from (\ref{htildeiexplicit}), which yields the second line of Eq. (\ref{hisolution}).

Let us examine the consequences of the requirement that for $h_i^{\rm dr}$ is real for $\lambda\in\mathcal{S, }$ as it was an implicit assumption that we are searching for purely real solutions of Eq. (\ref{dressing}).   The integral in the first line of (\ref{hisolution})\ will be real in the first interval in $\mathcal{S}$, namely in the interval $[\lambda_1,\lambda_2],$  if all the $a^{(j)}_i$'s are real.  The integral will give real values also on the consecutive intervals if and only if the following integral, \begin{align}
-\frac{2\imath}{\pi\hbar  c}\int_{\lambda_{2k}}^{\lambda_{2k+1}}\frac{\lambda^{i+1+g}+a^{(i)}_1 \lambda^{i+g}+a^{(i)}_2 \lambda^{i+g-1}+\dots+a_{i+g+1}^{(i)}}{\sqrt{\prod_{j=1}^{g+1}(\lambda-\lambda_j )}}\label{halfcycle},
\end{align}
is real for any $1\leq k\leq\ g$, but since the  $a^{(j)}_i$'s are real this requires the integral to vanish. In light of Eq. (\ref{dOmega0ExplBethe}) this is equivalent to the demand that all $a$-cycles of $d\Omega^{(Q)}_i$ are 0 (the integral in Eq.(\ref{halfcycle}) is just half the cycle), which uniquely fixes all the  $a^{(j)}_i$'s to the same values they take on the classical solution, Eq. (\ref{dOmega0Explicit}). 

We have thus defined  $\Omega_i^{(Q)}$ through the Bethe roots in the thermodynamic limit, Eq. (\ref{OmegaiFromDressing}),  and have shown that $\Omega_i^{(Q)}=\Omega_i^{(c)}$, which for $i=0$ ensures that the correspondence between classical and quantum states is based on both quantum and classical system agree on the values for the conserved quantities, as $\Omega_0$ serves as the generator of conserved quantities.

\section{Convergence of Generalized Hydrodynamics to the Whitham Equations}
We have  established, in the previous section, that solutions to the thermodynamic Bethe ansatz equations, Eq. (\ref{sigmas}), with $n$ given by the indicator function (characteristic function) of $\mathcal{S}=\cup_{i=1}^{g+1}[\lambda_{2i-1},\lambda_{2i}],$ give rise, in the $c\to0,$ limit to a root density given, up to a factor of    $\frac{\pi \imath c}{L}, $ by the jump discontinuity of the integral of a meromorphic differential $\Omega_0$ on the Riemann surface $y^2=\prod(\lambda-\lambda_i)$, and, since the root density on the one hand and the jump discontinuity of $\Omega_0$ serve as the generating function for conserved quantities on the quantum (Eq. (\ref{QIasFofRoots}))\ and classical sides (Eq. (\ref{IkClassical})), respectively, we have also shown that the quantum system (whose state is fixed by the Bethe roots) and the classical oscillatory solution (whose state is fixed by the Riemann surface) have the same conserved quantities. Note that we have dropped above, as we shall also do  below, the superscript $(c)$\ and $(Q)$ from $\Omega_i$ as whether the objects in question are classically defined or are the quantum analogues should  cause less confusion from now on. 

We now turn to the slow modulation of the classical oscillatory solutions and the quantum eigenstates as prescribed by the Whitham equations and generalized hydrodynamics, respectively, and show that  the two methods agree in the semiclassical limit, as should be the case, as both theories rely on the same principle of the slow modulation being completely constrained by the requirement of adhering to an infinite number of conservation laws.
\subsection{The Whitham Equations}
The Whitham equations may be summarized as follows:
\begin{align}
\partial_t \Omega_0=-\frac{2}{m}\partial_x \Omega_1,\label{Whitham}
\end{align}
which is equivalent to an infinite number of conservation laws (\ref{continuitydens}) written through the generating functions $\Omega_0$, $\Omega_1$ (Eqs. (\ref{IkThroughLogT}, \ref{JkThroughLogT}), respectively) for the appropriate integrated quantities (\ref{IJthroughpj}). We shall first want to motivate somewhat more the continuity equation (\ref{continuitydens}) or rather the fact that $\Omega_1$ is the generating function for  $J_k$ (or equivalently $j_k$), and  shall then turn to methods of solving these equations. 

Recall the Baker-Akhiezer function as being the function solving the linear problem associated with with Lax pair $U$ and $V$, Eq. (\ref{LinearProblems}). An explicit expression for the function will be given below, Eqs. (\ref{BA11explicit}, \ref{BA21explicit}), but here we shall only need the fact that the Baker-Akhiezer function has the form:
\begin{align} 
\Psi_{ij}(x,t;\lambda)=\varphi_{ij}(\bm A(\lambda)+\bm\phi(x,t))e^{\frac{\pm\imath}{\hbar}(\Omega_0x-\frac{2}{m}\Omega_1t) }\label{BAformGeneral}
\end{align}   
where $\varphi_{ij}$ are oscillatory functions, its argument $\bm A(\lambda)+\bm\phi(x,t)$ being an element of $\mathds{C}^g$.  The phase function $\bm \phi$ is  given by
\begin{align}
\phi_j=\oint_{b_j} \left(\Omega_0x-\frac{2}{m}\Omega_1t\right) 
\end{align}
while $\bm A(\lambda)$\ is the Abel map, a given function of the spectral parameter $\lambda$ given below. The point is then that averaging out over a period of the oscillations, one may compute
\begin{align}
\<\Psi^\dagger (-\imath \hbar \partial_x-\Omega_0)\Psi\>=0, \quad \left\<\Psi^\dagger \left(\imath \hbar \partial_t-\frac{2}{m}\Omega_1\right)\Psi\right\>=0\label{Averaged2}
\end{align} 
where $\Psi^\dagger$ is an appropriately defined dual to the Baker-Akhiezer function \cite{132:Krich:Intgr:AlgGeo,83:Krichever:Averaging} and $\<\dots\>$ denotes averaging over a period of the oscillations, which is quite intuitive given Eq.  (\ref{BAformGeneral}), as the oscillatory part averages out while $-\imath \hbar \partial_x$ and $\imath \hbar \partial_t$ pull $\Omega_0$ and $\frac{2}{m}\Omega_1$ from the exponent. Accepting this fact the compatibility conditions of the two equations in Eq. (\ref{Averaged2}), in analogy  to Eq. (\ref{LinearProblems}), becomes the Whitham equations, Eq. (\ref{Whitham}). Expanding the two function $\Omega_0,$\ and $\Omega_1$ around infinity yields coefficients that may be given the interpretation of conserved charges and currents, respectively,  since these may be written as polynomials of the spatial derivatives of the field variable of the non-linear Schr\"odinger equation, by using Eq. (\ref{Averaged2}) (writing, e.g., $\Omega_0=\frac{\<\Psi^\dagger (-\imath \hbar \partial_x)\Psi\>}{\<\Psi^\dagger \Psi\>}$) and taking account of the fact that the elements of the Baker-Akhiezer function, $\Psi$, contain the field variables on the off-diagonal (the elements $\Psi_{12}$ and $\Psi_{21}$ elements of the Baker-Akhiezer function).

Our goal in the previous paragraphs was only to motivate the fact that the conservation laws are encoded in the Whitham equations for the meromorphic differentials $\Omega_0$\ and $\Omega_1$, Eq. (\ref{Whitham}). A full and rigorous account of the procedure is more complicated and we refer the reader to the original papers\cite{83:Krichever:Averaging,132:Krich:Intgr:AlgGeo} and to reviews of this subject\cite{Carroll:Remarks:On:Whitham}, and proceed instead here to show how these equations may be integrated effectively\cite{132:Krich:Intgr:AlgGeo,83:Krichever:Averaging}. 

In order to integrate the Whitham equations, Eq. (\ref{Whitham}), one first notes that the equations  imply the existence of a potential $S,$ such that, \begin{align} 
\partial_{t_i}S=\Omega_i\label{action}
\end{align}
where $t_0=x$ and $t_1=-\frac{2}{m}$. In fact the Whitham equations generalize to any two pair of time $t_i$ and $t_j$ which are conjugate to an infinite set of Hamiltonians $I_k$ as follows\cite{83:Krichever:Averaging,132:Krich:Intgr:AlgGeo}
\begin{align}
\partial_k \Omega_j=\partial_j \Omega_k,
\end{align}
such that Eq (\ref{action}) holds for any pair of indices $i$, and $j$  both larger than $0$. The definition of $\Omega_i$ is given in Eq. (\ref{Omegaidef}). 

Within the Whitham approximation, the Baker-Akhiezer function takes the following form in terms of the action:
\begin{align}
\Psi_{ij}(x,t;\lambda)=\varphi_{ij}(\bm A(\lambda)+\bm\phi(x,t))e^{\frac{\pm\imath}{\hbar}S }
\end{align}
where 
\begin{align}
\phi_j=\oint _{b_j}dS.
\end{align}

We now discuss how a solution for $S$ may be found, as finding $S$ is equivalent to solving the Whitham equations. A solution for $S,$ or rather $dS\equiv\frac{dS}{dz}dz$ has the form:
\begin{align}
dS( \{t_k\})=\sum_{k=-1}^\infty({t_k}+t_k^{(0)})d\Omega_k(\{\lambda_i(\{t_l\}_{l=-1}^\infty)\}_{i=1}^{2g+2}),\label{KricheverSolution}
\end{align}
where $t_k^{(0)}$ are arbitrary and the branch points, $\lambda_i(\{t_k\})$, are to be chosen as functions of the times such that the singularity of $\frac{dS}{d\lambda} $ around each of the branch points, which is ostensibly of the form $\frac{1}{\sqrt{\lambda-\lambda_i}}$, in fact vanishes\cite{83:Krichever:Averaging}. That is the expansion around the branch point $\lambda_i$ is given by:
\begin{align}
\frac{dS}{d\lambda}   = O(\sqrt{\lambda-\lambda_i}).\label{KrichCondition}  
\end{align}

We shall show below that requiring  Eq. (\ref{KrichCondition}) of $S$\ automatically turns $S$ into a solution of the Whitham equations. This will be done  by following Ref. \cite{83:Krichever:Averaging}, but first let us re-formulate the solution to Eq. (\ref{action}) in order to hopefully achieve more clarity as how Eq. (\ref{KrichCondition}) can be used constructively to obtain a solution. Indeed, in order to find the times dependence of the branch points, given a set of initial conditions at times $\{t_k=0\}_{k=0}^\infty$, one first chooses  a set of $t_k^{(0)}$ such that if one constructs $dS$ according to  Eq. (\ref{KricheverSolution}) (in which one substitutes $t_k=0$), the singularity of $\frac{dS}{d\lambda}$ at the branch points vanishes according to Eq. (\ref{KrichCondition}). Then the evolution of the branch points at any later time is solved by finding such a position of the branch points so that if one constructs $dS $ again according to (\ref{KricheverSolution}), then the singularity at the branch points of $\frac{dS}{d\lambda}$ vanishes.

To show that Eq. (\ref{KrichCondition}) indeed provides a  solution of the Whitham equations, consider the singularities of  $\partial_{t_k}dS$ on the Riemann surface and compare that to the singularities of $d\Omega_k$. The behavior at $\infty_\pm$ of  $\partial_{t_k}dS$ is easily see to be the same as that of $d\Omega_k$, due to (\ref{KricheverSolution}). The differentials   $\partial_{t_k}dS$ could potentially have poles at the branch points since if $dS\sim d\lambda(\lambda-\lambda_i)^{-n/2}$ at such a points and then     $\partial_{t_k}dS \sim d\lambda(\lambda-\lambda_i)^{-n/2-1}$ which represents a pole of order $n$ in the local parameter $\sqrt{\lambda-\lambda_i}$, indeed $(\lambda-\lambda_i)^{-n/2-1}d\lambda=2(\lambda-\lambda_i)^{-(n+1)/2}d\sqrt{\lambda-\lambda_i} , $ but since we demand (\ref{KrichCondition}), $n=-1$ and    $\partial_{t_k}dS$  is analytic as a differential at the branch points. Lastly the $a_j$ cycles of $dS$ vanishes for any times due to the $a$-cycle condition in  (\ref{Omegaidef}). This means that     $\partial_{t_k}dS$  has all the same analytical properties of $d\Omega_k$, which determine it uniquely to be equal to $d\Omega_k$ and hence Eq. (\ref{action}) holds. 

 \subsection{Generalized Hydrodynamics and its Semiclassical Limit}
Generalized hydrodynamics is founded on a basis shared by Whitham theory, namely that the slow modulation of the state of the system may be predicted by the requirement of the continuity equations associated with an infinite number of conservation laws conserved Eq. (\ref{continuitydens}). 
The set $\{\rho_n\}_{n=1}^\infty$ fully describes any of the  translationally invariant  quantum states which are eigenstates of the Hamiltonian. The assumption then is that the non-equilibrium spatially varying quantum state may {\it locally} be described by such a translationally invariant state, the values of $\rho_n$ serving as space dependant  moduli of the state, to which Eq. (\ref{continuitydens}) serves as a set of equations of motion which dictate how these moduli depend on time. 

In order to apply this logic, the currents $j_n$ must be computed as a function of $\rho_n$ for any eigenstate of the Hamiltonian. Note that classically we already have an expression for $j_n$ since its generating function is given by $\Omega_1$, but we do not have yet the quantum expression for the generating function or for the individual $j_n$, we therefore give an argument for its derivation here,  a  rigorous proof was provided in Ref. \cite{Casto:Alvaredo:Doyon:Yoshimura:GHD}. 

To derive $j_n$ recall that an integrable system has an infinite number of times, $t_k$ conjugate to an infinite number of Hamiltonians, $\hat{I}_k$, each Hamiltonian being the integral over a Hamiltonian density $\int \hat{\rho}_k$. One of the times (in our case that would be $t_2$) while another time is designated as the spatial direction ($t_1$ in our case). The wavefunction is assumed to be periodic in the spatial direction. Imagine then choosing a new direction in which the wave function is assumed to be periodic and denote this perturbed  space variable  as  $t_1^{(\alpha)}.$  We take $t_1^{(\alpha)}=t_1^{(0)}+\alpha t_k^{(0)},$ where the $(0)$ superscript denotes unperturbed times.  The conjugate Hamiltonian densities associated with the perturbed space and time directions are then given by:
\begin{align}
\hat{\rho}^{(\alpha)}_1=\hat{\rho}^{(\alpha)}_1+\alpha \hat{\rho}^{(0)}_k\,\quad  \hat{\rho}^{(\alpha)}_2=\hat{\rho}^{(0)}_2\label{changeP}
\end{align}  
Note that the change of the momentum operator does not change the Hamiltonian and thus does not change the form of the Bethe ansatz as a linear superposition of plane waves with a given amplitude:
\begin{align}
\psi=\sum_{\sigma\in S_N} A(\sigma)e^{\imath k_i x_{\sigma(i)}}, \quad \frac{A(\tau^{(ij)} \sigma)}{A(\sigma)} = S(k_i,k_j),
\end{align}
where $\tau^{(ij)}$ is the transposition of the $i$-th and $j$-th variable and $S$ is the scattering matrix which is computed from knowledge of the Hamiltonian alone without reference to the momentum. 

We assume without proof that to leading order, the effect of the perturbation of the momentum operator is to change the quantization condition, namely it represents a change on the condition of periodicity of the wave function. As such we assume that it results in a change of the Bethe equation, which represents this periodicity condition, as follows:
\begin{align}
  e^{-\frac{\imath2 L}{\hbar} (\lambda_i +\alpha  \lambda_i^k)}\prod_{j= 1}^N\frac{\lambda_i-\lambda_j+\imath c/2}{\lambda_i-\lambda_j-\imath c/2}=-1.\label{perturbedBethe}
\end{align}

The result of the perturbation may also be seen by considering the force on the particles which is given by the time derivative of the momentum density. One obtains:
\begin{align}
\partial_t \hat{\rho}^{(\alpha)}_1=\partial_t\hat{\rho}_1^{(0)}+\alpha \partial_t\hat{\rho}_k^{(0)}=\partial_t\hat{\rho}^{(0)}_1+\alpha \partial_x\hat{j}^{(0)}_k,
\end{align}
here $t=t_2=t_2^{(0)}=t_2^{(\alpha)}$ is the usual unperturbed physical time variable.

To proceed we wish to apply the perturbation by introducing a perturbation to the Hamiltonian rather than to the momentum, which nevertheless induces the same force on the particles. We do this by perturbing $\hat{\rho}_2$ which serves as the Hamiltonian density (up to a  constant) as follows:
\begin{align}\hat{\rho}^{(\alpha)}_2=\hat{\rho}^{(\alpha)}_2+m\alpha \hat{j}^{(0)}_k\,\quad  \hat{\rho}^{(\alpha)}_1=\hat{\rho}^{(0)}_1\label{changeH}\end{align} 
Indeed, computing again the force operator  on the particles we have:
\begin{align}
\partial_t \hat{\rho}^{(\alpha)}_1(x)=\frac{\imath4}{m\hbar^2c}\oint [\hat{\rho}_2^{(0)}(y)+m\alpha\hat{j}_k^{(0)}(y),\hat{\rho}^{(0)}_1(x)]dy=\partial_t\hat{\rho}^{(0)}_1(x)+\alpha \partial_x\hat{j}^{(0)}_k(x),\label{TotalForcer}
\end{align}
the last term on the right hand side stems from the fact that $\hat{\rho}_1^{(0)}=-\frac{\imath\hbar^2c}{4}  \hat{\psi}^\dagger\partial_x \hat{\psi}$ such as its effect  when commuted with any integral over a local operator composed of the bosonic operators is to present the  derivative of the integrand times $\frac{\imath\hbar^2c}{4}$ as the result of the calculation. 

Assuming that  the scheme described in Eq. (\ref{changeP}) and the scheme described in Eq. (\ref{changeH}) are the same, based on the fact that the produce the same force on the particles, we may consider the change in energy $\delta \mathcal{E}$ due to the perturbation in both schemes and assume it to be also the same. First, note that to find the energy of the system under the scheme (\ref{changeP}) we must solve (\ref{perturbedBethe}) and then simply calculate the energy according to $\mathcal{E}=\sum E(\theta_i)$, where $E(\theta_i)=\frac{2\hbar^2 \theta_i^2}{m}$ This is a tractable problem in the thermodynamic limit thanks to the thermodynamic Bethe ansatz, thus a computation of $\frac{\delta \mathcal{E}}{\delta \alpha}$ is tractable as well, and shall be pursued in the next paragraphs. The significance of all this is that, by the scheme of Eq. (\ref{changeH}), and since $\hat{H}^{(\alpha)}=\frac{4}{m\hbar c}\int \hat{\rho}^{(\alpha)}_2=\hat{H}^{(0)}+\frac{4\alpha}{\hbar c}\int \hat{j}_k^{(0)},$  Hellman-Feynman dictates:
\begin{align}
j_k\equiv \<\hat{j}_k\>=\frac{\hbar c}{4} \frac{\delta \mathcal{E}}{\delta \alpha}, 
\end{align}   
that is, we have presented a prescription to calculate the currents. 

To solve Eq. (\ref{perturbedBethe}) we proceed along similar lines to those that produced Eq. (\ref{SigmasDressing})\ from Eq. (\ref{BetheEqs}). The logarithm of Eq. (\ref{perturbedBethe}) reads:
\begin{align}
(j_i+1/2)=\frac{1}{\pi\hbar }\left(\theta^{(\alpha)}_i +\alpha (\theta_i^{(\alpha)})^k\right)+\frac{1}{2\pi L}\imath\sum_j  \log\frac{\theta^{(\alpha)}_i-\theta^{(\alpha)}_j+\imath c/2}{\theta^{(\alpha)}_i-\theta^{(\alpha)}_j-\imath c}.  
\end{align}
Writing $\theta_i^{(\alpha)}=\theta_i+\delta\theta_i,$ where $\theta_i$ solve the Bethe equations for $\alpha=0$, one obtains, to first order in $\alpha,$
\begin{align}
0=\frac{1}{\pi\hbar }\left(\delta\theta_i +\alpha \theta_i^k\right)+\frac{c}{2\pi L}\sum_j  \frac{\delta\theta_i-\delta\theta_j}{(\theta_i-\theta_j )^2+ c^2/4}  .\label{perturbedRelation}  
\end{align}
We may Pass to the continuum limit by defining a function  $\delta \lambda(\lambda)$ which is a function derived from $\delta \theta_i$\ by interpolating the relation  $\delta\lambda(\theta_i)\equiv\delta\theta_i$.  Then using this function Eq. (\ref{perturbedRelation}) can be written in the continuum limit as:

\begin{align}
\sigma_s(\lambda)\delta\lambda(\lambda)=-\frac{\alpha \lambda^ k}{\pi\hbar }+\int \frac{c}{(\lambda-\lambda')^2+ c^2/4}  n(\lambda')\sigma_s(\lambda')\delta\lambda(\lambda)\frac{d\lambda'}{2\pi}, 
\end{align} 
where  Eq. (\ref{sigmas}) was used to simplify the resulting expression. 

We thus have
\begin{align}
\sigma_s(\lambda)\delta\lambda(\lambda)=-\alpha h_k^{\rm dr}
\end{align}
The change in energy is equal to $\delta \mathcal{E}=\sum_i \delta \lambda_i \frac{dE(\lambda_i)}{d\lambda_i}$, which  in the continuum limit  becomes  $\delta \mathcal{E}=\int n \sigma_s\delta\lambda \frac{dE(\lambda)}{d\lambda}$, from which we obtain:
\begin{align}
j_k=\frac{\hbar c}{4}\frac{\delta \mathcal{E}}{\delta \alpha}=-\frac{\hbar c}{4}\int nh_k^{\rm dr} dE=-\frac{\hbar c}{m}\int nh_1^{\rm dr} \lambda^k d\lambda\label{h1generatej} 
\end{align} 
which shows that the $nh_1^{\rm dr}$ is the generating function for the currents. The last equality stems from the Hermiticity of the dressing operator when followed by a multiplication by $n$. Namely $\int a  n b^{\rm dr}=\int a^{\rm dr}nb$. The Hermiticity of the dressing operation times $n$ easily follows from its representation as $na^{\rm dr}=n(id-\phi\star n)^{-1}a,$ where $\phi\star$ is the operation of convolution with $\frac{2c}{\lambda^2+c^2},$ and $id$ is the identity operator. The operator $n(id-\phi\star n)^{-1}$ is hermitian under the integral scalar product $a\cdot b=\int ab$ by straightforward operator analysis.
 
Generalized hydrodynamics then follows form the condition $\partial_t \rho_k=\partial_x j_k$, for each $k$. We have just established that the generating function for the currents is  $nh_1^{\rm dr}$ (Eq. (\ref{h1generatej})). The generating function for the densities is $nh_0^{\rm dr}$ follows form (\ref{QIasFofRoots}):
\begin{align}
\rho_k=\frac{\hbar c}{2}\int n\sigma_s\lambda^k=\frac{\hbar c}{2}\int h_0^{\rm dr}\lambda^k,
\end{align}  which suggests that continuity condition in terms of generating functions is given by:
\begin{align} 
-\frac{2}{m}\partial_x (nh_1^{\rm dr})= \partial_t (nh_0^{\rm dr}).
\end{align}
We are already familiar with the semiclassical limit of $nh_k^{\rm dr}$. These are the jump discontinuity of the  $\frac{\hbar c}{4\pi\imath }\Omega_k$'s (Eq. (\ref{OmegaiFromDressing})), respectively, such that, in the semiclassical limit, generalized hydrodynamics may be written as:
\begin{align} 
-\frac{2}{m}\partial_x (\Omega_1(\lambda+\imath0^+)-\Omega_1(\lambda+\imath0^-))=\partial_t (\Omega_0(\lambda+\imath0^+)-\Omega_0(\lambda+\imath0^-)),
\end{align}
for real $\lambda$. This relation shows that the jump discontinuity across the real axis of $\partial_x\Omega_1$ is equal to the jump discontinuity across the real axis of $\partial_t\Omega_0, $ but since both left and right hand sides are a-priori analytic functions away from the cuts, and thus determined uniquely by their jump discontinuity across the real axis, the latter equation also leads to a more straightforward relation between $\Omega_1$ and $\Omega_0$ being given just by the Whitham equations Eq. (\ref{Whitham}).  Thus we have shown that generalized hydrodynamics converges to Whitham theory in the semiclassical limit.

\section{Convergence of quantum matrix elements to classically averaged quantities}
In this section we shall first recount the relation between classical averages and quantum matrix elements as discovered by Babelon Bernard and Smirnov \cite{Smirnov:Babelon:Defoned:Hyperlliptic}. This relation is based on the expression of classical averages over the period oscillations in separated variables due to Flaschka, Forest and McLaughlin \cite{88:Flaschka:KdV:Avging} and application of the inverse scattering method to write quantum matrix elements in terms of separated variables based on the work of Sklyanin (See Ref.\cite{Sklyanin:SoV} and references therein). We use this method to compute the momentum average at points close to a topological transition of the algebraic Riemann surface (an annihilation of a branch cut or a creation of a branch cut) describing the modulated Whitham flows that characterize the quantum Lieb-Liniger model at small $c$, namely at the semi-classical limit. The momentum average shows characteristic oscillations similar to those found at infinite $c$, namely for free fermions, in Ref. \cite{Bettelheim:Glazman}.\ 
\subsection{Averages over Classical Oscillating Solutions}
In order to calculate average quantities over the oscillating solutions we first derive explicit equations for the Baker-Akhiezer function of a  general multi-phase solution, namely solutions that correspond to a any given hyper-elliptic Riemann surface with moduli $\{\lambda_i\}_{i=1}^{2g+2}$.  To be able to do so, some preliminary definitions must be first established. 

 We define holomorphic differentials $d\omega_i$, where $i=1,2,\dots,g$, having the form
\begin{align}
d\omega_i=\frac{a_1^{(i)}\lambda^{g-1}+a_2^{(i)}\lambda^{g-2}+\dots+a_g^{(i)}}{\sqrt{\prod(\lambda-\lambda_i)}} d\lambda, 
\end{align}
where the coefficients $a_j^{(i)}$ are fixed by the conditions:
\begin{align}
\oint_{a_j} d\omega_i=2\pi\imath\delta_{ij},
\end{align}
as a result the coefficients $a_j^{(i)}$ are all imaginary. The matrix of $b$-periods, $\bm{B},$ can then be constructed as follows:
\begin{align}
B_{ij}= \oint_{b_j} d\omega_i.
\end{align}
This matrix then has all real coefficients. 
The Riemann theta function can then be constructed:
\begin{align}
\theta(\bm{v})=\sum_{\bm{m}\in\mathds{Z}^g} e^{\frac{1}{2}\bm{m}^t\bm{B} \bm{m} + \bm{m}^t\bm{v}}.\label{RiemannTheta}
\end{align}
Since $\bm{B}$ is real $\bar\theta(\bm{v})=\theta(\bm{v})$. 

The Abel map is defined by
\begin{align}
\bm{A}(\lambda)=\left(\begin{array}{c}\int_{\infty_-}^\lambda d\omega_1 \\ \int_{\infty_-}^\lambda d\omega_2\\.\\.\\.\\\int_{\infty_-}^z d\omega_g \end{array} \right).
\end{align}
The map is an homomorphism of the Riemann surface onto a two dimensional surface in $\mathds{C}^g/(2\pi\imath \mathds{Z}^g+\bm{B} \mathds{Z}^g).$ The map is real on the real axis for points outside the cuts on the upper sheet modulo $2\pi\imath \bm{m}+\bm{B}\bm{n}$ for  $\bm{n},\bm{m}\in \mathds{Z}^g. $ Note that the Riemann theta function, Eq. (\ref{RiemannTheta}) is quasi-periodic on this  $\theta$  in $\mathds{C}^g/(2\pi\imath \mathds{Z}^g+\bm{B} \mathds{Z}^g)$:
\begin{align}
\theta(\bm{v}+2\pi\imath\bm{m}+\bm{B}\bm{n})=e^{\bm{n}^t\bm{v}}\theta(\bm{v})=\sum_{\bm{m}\in\mathds{Z}^g} e^{\frac{1}{2}\bm{m}^t\bm{B} \bm{m} + \bm{m}^t\bm{v}},
\end{align}
where $\bm{m},\bm{n}\in \mathds{Z}^g$. The importance of the Riemann theta function is that any meromoprhic function on the Riemann surface can be expressed through  combinations of this function.

Choose $g$ points, $\lambda_1,$ $\lambda_2,$ $\dots,$ $\lambda_g$  on the Riemann surface and define \begin{align}
\bm{D}=\sum_{i=1}^g \bm{A}(\gamma_i).
\end{align}
Further we repeat the definition of the meromorphic differentials, $d\Omega_{i}$, with $i\geq-1$,  as having singularities at $\infty_\pm$ of the form:
\begin{align}
d\Omega_{i}\sim \pm d\lambda \lambda^{i}
\end{align}
and normalized by the condition        
\begin{align}
\oint_{a_j} d\Omega_i =0,
\end{align}
leading further to the definition of the functions $\Omega_i$ by:
\begin{align}
\Omega_i(z) =\int_{\lambda_1}^z d\Omega_i
\end{align}
These functions are real on the real axis outside of the cuts on the upper sheet. They are multi-valued function, so may be only treated as functions if their value is taken modulo $\bm m \bm{W}_i$, with   
\begin{align}
\bm{W}_i=\left(\begin{array}{c}\oint_{b_1} d\Omega_i\\\oint_{b_2} d\Omega_i\\.\\.\\.\\\oint_{b_g} d\Omega_i \end{array}\right)
\end{align} 
and $\bm{m}\in \mathds{Z}^g$.

The functions $\Omega_i$  have the following asymptotics on the real axis with 
\begin{align}
\Omega_0\sim \pm\left(\lambda-\frac{p_0}{2} \right), \quad \Omega_1\sim \pm\frac{1}{2}\left(\lambda^2-\frac{mE_0}{2} \right),\quad \Omega_{-1}\sim \pm\log\left(\frac{\lambda}{\beta}\right)
\end{align}
as $\lambda \to \infty_\pm$, respectively, where $p_0$, $E_0$ and $\beta$ are all real. Define also 
\begin{align}
\bm{P}=\frac{1}{\hbar}\bm{W}_0, \quad \bm{E}=-\frac{1}{2m\hbar}\bm{W}_1, \quad \bm{r}=\left(\begin{array}{c}\oint_{b_1} d\Omega_{-1}\\\oint_{b_2} d\Omega_{-1}\\.\\.\\.\\\oint_{b_g} d\Omega{-1} \end{array}\right)=-\bm{A}(\infty_+)
\end{align} 
The vectors $\bm{P}, $ $\bm{E}$, $\bm{r}$ are all purely imaginary. 

The Baker-Akhiezer function has the following expression:
\begin{align}
&\Psi_{11}(z)=\frac{\theta(\bm{A}(z)+\imath \bm{P} x +\imath\bm{E} t-\bm{D})\theta(\bm{D})}{\theta(\boldsymbol{A}(z)-\bm{D})\theta(\imath \bm{P} x +\imath\bm{E} t-\bm{D})}e^{\frac{ \imath x}{\hbar}( \Omega_0(z)+p_0/2)-\frac{\imath2 t}{m\hbar}( \Omega_1+mE_0/2)}\label{BA11explicit}\\
&\Psi_{21}(z)=\alpha\beta\frac{\theta(\bm{A}(z)+\imath \bm{P} x +\imath\bm{E} t+\bm{r}-\bm{D})\theta(\bm{D}+\bm{r})}{\theta(\boldsymbol{A}(z)-\bm{D})\theta(\imath \bm{P} x +\imath\bm{E}t-\bm{D})}e^{\frac{ \imath x}{\hbar}( \Omega_0(z)-p_0/2)-\frac{\imath2 t}{m\hbar} (\Omega_1-mE_0/4)+\Omega_{-1}}\label{BA21explicit}
\end{align}
while the other elements of the matrix are given by making use of the following:
\begin{align}
\Psi_{i2}(z)=\Psi_{i1}(\sigma(z)),\label{BAInvolution}
\end{align}
where $\sigma$ is an operator that switches the sheets. Namely, $\sigma(z_\pm)=z_\mp.$ The explicit expressions above for the Baker-Akhiezer function allows one to compute the monodromy matrix as well, as done in the appendix. The explicit expressions for the matrix elements of the Baker-Akhiezer function also allows one to identify $\Phi$ in   (\ref{BAasymptotics})
\begin{align}
\bm{\Phi}^{(1)}=\left(\begin{array}{cc} \Phi^{(1)}_{11} & \frac{\theta(\imath \bm{P} x +\imath\bm{E} t-\bm{D}-\bm{r})\theta(\bm{D})}{\alpha \theta(\bm{D}+\bm{r})\theta(\imath \bm{P} x +\imath\bm{E} t-\bm{D})}  e^{\frac{\imath}{\hbar} (x p_0- t E_0) } \\ \alpha\beta^2\frac{\theta(\imath \bm{P} x +\imath\bm{E} t-\bm{D}+\bm{r})\theta(\bm{D}+\bm{r})}{\theta(\bm{D})\theta(\imath \bm{P} x +\imath\bm{E}t-\bm{D})}e^{-\frac{\imath}{\hbar} (x p_0- t E_0) } & \Phi^{(1)}_{22} \end{array} \right)
\end{align}
The parameter $\alpha$ can be chosen such that $\psi$ is indeed the complex conjugate of $\bar{\psi}$, assuming that $\gamma_i$ are chosen on the real axis. This choice is given by:
\begin{align}
\alpha=\frac{\theta(\bm{D})}{  \beta|\theta(\bm{D}+\bm{r})|}\label{alphaExpression}  
\end{align}
such that $\psi$ is given by:
\begin{align}
\psi=  \frac{\theta(\imath \bm{P} x +\imath\bm{E} t-\bm{D}-\bm{r})}{ \theta(\imath \bm{P} x +\imath\bm{E} t-\bm{D})}  e^{\imath\left(\frac{\pi}{2}-\arg \theta(\bm{D}+\bm{r})+\frac{x p_0- t E_0}{\hbar} \right) +\log(\beta)}
\end{align}

To perform the averages it is useful to write the expression for $\psi$, which involves theta functions, in terms of the zeros of the theta function. To that end consider the zeros of $F(\bm{A})=\theta(\bm{A}+\imath \bm{P} x +\imath\bm{E} t-\bm{D})$ denoted $\mu_1, $ $\mu_2,$ $\dots,$ $\mu_g$. Their sum   may be computed as follows:
\begin{align}
\sum_{i=1}^g \mu_i=-\frac{1}{2\pi\imath}\oint z \frac{dF(\bm{A}(z))}{F(\bm{A}(z))},
\end{align}
where the integral is to be taken around two cycles  covering  infinity on both sheets, respectively,  both going counterclockwise. By expanding $F$\ around infinity one obtains:
\begin{align}
&\sum_{i=1}^g \mu_i=-\frac{1}{2\pi\imath}\sum_i\oint \frac{a_1^{(i)}z^{g-1}+a_2^{(i)}z^{g-2}+\dots+a_g}{\sqrt{R}} z \partial_i\log(F)dz=\sum_{\pm,i} \pm a_1^{(i)} \partial_i\log(F(\bm{A}))|_{\bm{A}=\bm{A}(\infty_\pm)}\nonumber=\\&=\sum \bm{P}\cdot\bm{\nabla}\log \frac{\theta(\imath \bm{P} x +\imath\bm{E} t-\bm{D}-\bm{r})}{\theta(\imath \bm{P} x +\imath\bm{E} t-\bm{D})}=-\imath\sum\partial_x\log \frac{\theta(\imath \bm{P} x +\imath\bm{E} t-\bm{D}-\bm{r})}{\theta(\imath \bm{P} x +\imath\bm{E} t-\bm{D})}
\end{align}
which allows us to write:
\begin{align}
\frac{\theta(\imath \bm{P} x +\imath\bm{E} t-\bm{D}-\bm{r})}{\theta(\imath \bm{P} x +\imath\bm{E} t-\bm{D})}=e^{-\imath\sum_i\int \mu_i(x,t) dx} 
\end{align}
Thus the field in the nonlinear Schr\"{o}dinger equation takes the form:
\begin{align}
\psi(x)=\beta e^{\imath\int_{}^{x} \left[p_0-\sum_i\mu_i(x')\right]dx'-\imath \int^{t}E_0dt' }.\label{psiAsMus}
\end{align}
We have translated the formula for the field variable $\psi$ to a a formula which depends on the variables $\mu_i$ which have periodic orbits between the $\lambda_i$. From this expression, any combination of field variables and their derivatives may be written through the $\mu_i$'s. 

The prescription for averaging over the oscillatory solution was arrived to in Ref. \cite{88:Flaschka:KdV:Avging}  to be given by an integral over the cycles of the $\mu_i$'s with the measure $\frac{\prod_{k<j}(\mu_k-\mu_j)\prod d\mu_j}{\prod_j\sqrt{R(\mu_j)}} $ with proper normalization. More explicitly:
\begin{align}
&\<f(\{\mu_i\})\>\equiv\frac{1}{L}\int _{x}^{x+L}f(\{\mu_i\})=\frac{\oint f(\{\mu_i\})\frac{\prod_{k<j}(\mu_k-\mu_j)\prod d\mu_j}{\prod_j\sqrt{R(\mu_j)}}}{\oint\frac{\prod_{k<j}(\mu_k-\mu_j)\prod d\mu_j}{\prod_j\sqrt{R(\mu_j)}}}\label{classicalavging}
\end{align}

\subsection{The Bethe Wavefunction in Separated Coordinates}
In the quantum case the monodromy matrix $\hat{\bm{T}}$ becomes a matrix with operator matrix elements. The element $\hat B$ (namely $\hat T_{12}$ ) commutes for different values of the spectral parameter
\begin{align}
[\hat B(\lambda_1),\hat B(\lambda_2)]=0.
\end{align}
which allows to denote a set of commuting operatos $\hat \gamma_i$ which act as the zeros of   $B$ , namely 
\begin{align}\hat B(\lambda)=\prod_{i}(\lambda-\hat{\gamma}_i).\end{align} One has the following algebra
\begin{align}
&\hat A(\mu)\hat B(\lambda)=\left(\frac{\lambda-\mu-\imath c/2}{\lambda-\mu}\right)\hat B(\lambda)\hat A(\mu)+\frac{\imath c/2}{\lambda-\mu}\hat B(\mu)\hat A(\lambda )\\
&\hat B(\lambda)\hat A(\mu)=\left(\frac{\lambda-\mu+\imath c/2}{\lambda-\mu}\right)\hat A(\mu)\hat B(\lambda)-\frac{\imath c/2}{\lambda-\mu}\hat A(\lambda )\hat B(\mu)
\end{align}  
We may choose a basis in which $\hat{\gamma}_i$ are all diagonalized. In such a basis $\hat{\gamma}_i$ can be treated as $c$-numbers and $B$ as a $c$-function depending on $\gamma_i$. The above algebra,  complemented by the algebra of $\hat{D}$ with $\hat{B}$ which we do not show here, suggests that  $\hat{A}(\gamma_i)$ and $\hat{D}(\gamma_i)$ may be written as operators on the space of functions of $\gamma_i$. The operators consistent with the algebra above and with the behavior at large $\gamma_i$ may be written as\cite{Sklyanin:SoV}:
\begin{align}
\hat{A}(\gamma_i) =\epsilon_i e^{\imath \gamma_i L/2\hbar }e^{-\imath c/2\partial_{\gamma_i}}, \quad \hat{D}(\gamma_i)=\tilde \epsilon_i e^{-\imath \gamma_i L/2\hbar }e^{\imath c/2\partial_{\gamma_i}},
\end{align}
From $\hat{\bm{T}}^*=\sigma_y \hat{\bm{T}}\sigma_y$ one obtains $\tilde \epsilon_i=\epsilon_i$ and from $\det_q(\bm{T})=e^{-cL/4\hbar }$ one obtains $\epsilon_i=\pm e^{-cL/8\hbar}$. Here the quantum determinant is given by\begin{align}
\det\,_q(\hat{\bm{T}})\equiv\hat A(\lambda)\hat D(\lambda+\imath c/2)-\hat B(\lambda)\hat{C}(\lambda+\imath c/2),
\end{align}
and represents an object lying in the center of the algebra of the matrix elemetns of $\hat{\bm{T}}$.
Choosing $\lambda=\gamma_i$ one obtains:
\begin{align}
&\det\,_q(\bm{T})=\hat A(\gamma_i)\hat D(\gamma_i+\imath c/2)-\hat B(\gamma_i)\hat{C}(\gamma_i+\imath c/2)=\epsilon_i e^{\imath \gamma_i L/2\hbar }e^{-\imath c/2\partial_{\gamma_i}} \hat D(\gamma_i+\imath c/2)=\\&=\epsilon e^{\imath \gamma_i L/2\hbar }\hat D(\gamma_i) e^{-\imath c/2\partial_{\gamma_i}}=\epsilon_i^2     = e^{-cL/4\hbar } 
\end{align}

The eigenfunction of $\tr(\hat{\bm T}(\lambda))$ for $\lambda=\gamma_i$ then is some wavefunction of $\{\gamma_j\}_{j=1}^N$, denoted by $\psi(\{\gamma_j\}_{j=1}^N)$ satisfying the eigenvalue equation:
\begin{align}
\tr(\hat{\bm T}(\gamma_i))\psi(\{\gamma_j\}_{j=1}^N)=\epsilon_i \left(e^{\imath \gamma_i L/2\hbar }e^{-\imath c/2\partial_{\gamma_i}}+e^{-\imath \gamma_i L/2\hbar }e^{\imath c/2\partial_{\gamma_i}}\right)\psi(\{\gamma_j\}_{j=1}^N)=t(\gamma_i)\psi(\{\gamma_j\}_{j=1}^N)\label{TQprecursor}
\end{align}
In this basis there is no interaction between the different variables $\gamma_j$ and so the wavefunction separates $\psi(\{\gamma_j\}_{j=1}^N)=\prod_{j=1}^N e^{\frac{\pi(4 n_j+{1-\epsilon_i})\gamma_i}{c}}Q(\gamma_j), $ where $n_j$ is some integer to be determined later and each wavefunction obeys the $T-Q$\ equation as a direct consequence of (\ref{TQprecursor}):
\begin{align}
 e^{cL/4\hbar }t(\gamma)Q(\gamma)=e^{\frac{-\imath\gamma L}{2\hbar }}Q(\gamma+\imath c/2)+e^{\frac{\imath\gamma L}{2\hbar }}Q(\gamma-\imath c/2).\label{TQ}
\end{align}

We shall solve this equation in the semiclassical limit following Babelon, Bernard, Smirnov  \cite{Smirnov:Babelon:Bernard:Null:Vectors,Smirnov:Babelon:Bernard:Qauntization:Solitons,Smirnov:Babelon:Defoned:Hyperlliptic} and Smirnov \cite{Smirnov:Quasi-Classical:qKdV} in order to obtain matrix elements and expectation values in that limit in the next subsection.

Equation (\ref{TQ}) leads to the Destri de-Vega equation:
\begin{align}
\log \frac{a(\gamma)}{a(\gamma-\imath c/2)}=\frac{1}{2\pi\imath}\oint \log \left(\frac{\gamma-\mu+\imath c/2}{\gamma-\mu-\imath c/2}\right) d\log(a(\mu)+1)\label{DestriDeVega}
\end{align}
for $a(\mu)\equiv \frac{Q(\mu+\imath c/2)}{Q(\mu-\imath c/2)}$which is an expression of the fact that $a(\lambda )+1=0$ (producing on the right hand side a log pole at $\lambda+\imath c/2$ and a log zero $\lambda-\imath c/2$) implies $Q(\lambda)=0$ which in turn implies that $\log a(\lambda)$ has a log  zero at $\lambda-\imath c/2$ and a log pole at $\lambda+\imath c/2,$ while $a(\lambda)+1=\infty$ (producing on right hand side a log zero at $\lambda+\imath c/2$ and a log pole $\lambda-\imath c/2)$ implies  $Q(\lambda-\imath c/2)=0$ which implies $\log a(\mu-\imath c/2)$ has a log zero at $\lambda-
\imath c/2$ and a log pole at $\lambda+\imath c/2.$ The equation may be solved to leading order in $c$ by taking $a(\mu)=e^{2L\imath\Omega_0/\hbar }$ . Indeed, writing $\log(a(\mu)+1)=\log(2\cos(L\Omega_0/\hbar ))+\imath L \Omega_0\hbar =\log(t)+\imath L\Omega_0 /\hbar ,$ and taking into account that $t$ has no zeros or poles in the first sheet,  gives for the integral in (\ref{DestriDeVega}) the following:
\begin{align}
\frac{1}{2\pi\imath}\oint \log \left(\frac{\gamma-\mu+\imath c/2}{\gamma-\mu-\imath c/2}\right) d\log(a(\mu)+1)=-\frac{\imath  L}{\hbar \pi}\oint  \frac{c/2}{(\gamma-\mu)^2+c^2/4}d\Omega_0\overset{c\to0}\to-\frac{Lc\partial \Omega_0(\gamma)}{\hbar } ,
\end{align}
which is the limit of the left hand side of (\ref{DestriDeVega})\ as well, as required.

Having the leading order solution one may continue to the subleading order, which is necessary since the leading order is not sufficient to the required accuracy. We continue to follow in this section mainly Smirnov\cite{Smirnov:Quasi-Classical:qKdV}, making only very minor changes related to the fact that we are dealing with the nonlinear Schr\"odinger equation, rather than then Korteweg-de Vries equation.
One writes,  $a=e^{\frac{2L\imath \Omega_0}{c\hbar }}(1+\frac{c x}{2} )$ to obtain the following equation:
\begin{align}
  x(\gamma)+\frac{L}{\hbar} \partial\Omega_0(\gamma)=\frac{1}{2\pi\imath }\oint \frac{1}{\mu-\gamma} \frac{e^{L\imath \Omega_0(\mu)/\hbar }x(\mu)}{\cos(L\Omega_0(\mu)/\hbar )}d\mu. 
\end{align}
One may deform the contour to surround the zeros of $\cos(L\Omega_0/\hbar )$ picking up the pole at $\mu=\gamma$ to obtain:  
\begin{align}
\mathcal{P}_-\tan(L\Omega_0(\gamma)/\hbar)x(\gamma)=\frac{L}{\hbar}    \Omega_0(\gamma),
\end{align}
where $\mathcal{P}_-$ projects a meromorphic function to a function which has the same singularities on the lower sheet but no singularities on the upper sheet.
To this equation one should add that $x(\gamma)d\gamma$\ is regular on the upper sheet. One then obtains:
\begin{align}
x(\gamma)d\gamma=\mathcal{P}_-d\log\sin (L \Omega_0/\hbar).\label{xasprojected}
\end{align} 
which may be  written as an infinite sum:
\begin{align}
x(\gamma)d\gamma= \sum_{\Omega_0(\mu_i)=\frac{i\pi}{L}}d\Omega^-_{\mu_{i}}(\gamma),
\end{align}
where$ d\Omega^-_{\mu}$ is a differential which has a pole of residue $1$ at $\mu$ and a pole of residue $-1$ at infinity both on the lower sheet and has a vanishing $b$-cycles. Another important consequence of Eq. (\ref{xasprojected}) is the following:
 \begin{align}
(x(\gamma)+x(\sigma \gamma))d\gamma=d \log\sin \frac{L\Omega_0(\gamma)}{\hbar}   ,\label{xproperty}
\end{align}
Given that $a(\mu)\equiv \frac{Q(\mu+\imath c/2)}{Q(\mu-\imath c/2)}$ one may write:
\begin{align}
Q=e^{-\imath \int \left(\frac{2L\Omega_0}{c\hbar } +x\right) d\gamma }\label{Qresult} 
\end{align}
which, when combined with (\ref{xproperty}), gives:
\begin{align}
Q(\gamma)\bar Q(\sigma \gamma)=\frac{e^{-\frac{ 4\imath }{\hbar c}\int ^\gamma L\Omega_0(\gamma')d\gamma' }}{\sin(L\Omega_0(\gamma)/\hbar )},
\end{align}
which is the key property of $Q$\ which we shall need in the sequel, since it represents the wavefunction multiplied by its adjoint, or rather an important piece of which. In fact, taking into account the relation \begin{align}\chi(\{\gamma_j\}_{j=1}^N)=\prod_{j=1}^N e^{\frac{\pi(4 n_j+{1-\epsilon_i})\gamma_i}{\hbar c}}Q(\gamma_j), \label{QchiRelation}\end{align} one may write for the wave function:
\begin{align} 
\chi^+\chi=\prod_j \frac{e^{-\frac{ 4\imath }{ c}\int ^{\gamma_j}\left( L\Omega_0/\hbar -\pi\imath j\right)d\gamma }}{\sin(L\Omega_0(\gamma_j)/\hbar)},\label{normsimple}
\end{align}   
where we assumed $n_j +\frac{1-\epsilon_j}{4}=\frac{j}{2}$ namely we assume that all mode numbers are represented twice in the set with repetitions $\{n_j\}_{j=-\infty}^\infty$ such that $n_j=\left\lfloor \frac{j}{2}\right\rfloor$ can be chosen and we assume further $\epsilon_j=(-)^j$. This choice is necessary in order for the semiclassical limit to agree with classics.

In order to use (\ref{normsimple}), we shall need more  precise  information on $\Omega_0,$  than that provided in the approximation of $\Omega_0$ as  $h^{\rm dr}_0$ (Eq. (\ref{Omega0FromDressing})). In fact we shall demand Eq. (\ref{demandModeNumber}) below. To see the origin of this requirement , we first recall the relation between $\Omega_0$ and $\Lambda$, where the latter is defined in (\ref{LambdaDef}).  
The logarithm of the Bethe equations (\ref{logBetheEqs}) take the following form
in terms of $\Lambda$:
\begin{align}
g(\lambda_i)=-\pi\imath(2 j_i+1),\quad g(\lambda)\equiv\log\left[ e^{\frac{cL}{\hbar 2}}\Lambda(\lambda)\Lambda(\lambda+\imath c/2)\right].
\end{align}
Given the fact that as $c\to0$, we have $\Lambda\to e^{\frac{\imath}{\hbar} L\Omega_0},$ Eq. (\ref{LambdaLimit}), we can identify -$\frac{L}{\pi\hbar   }\Omega_0+\frac{\imath cL}{4\hbar\pi}       $ as the mode number $j_i$ of the Bethe root. We demand that the mode number changes by an integer over any gap:\begin{align}\frac{L}{2\pi\hbar }\oint_{a_j} d\Omega_0\in\mathds{Z,}\label{demandModeNumber}\end{align}such that at the edge of each gap there will be an available state for a Bethe root to occupy. This requirement is consistent with the fact that $\oint_{a_j}d\Omega_0$ is already determined by the normalization $\oint_{a_j}d\Omega_0=0$ since the mode number requirement can be satisfied by adding a $\frac{1}{L}$ correction to the classically determined value of $\oint_{a_j}d\Omega_0$.

Eq. (\ref{normsimple}) features the object $\sin(L\Omega_0/\hbar)$  which can be represented in a more convenient form to facilitate developing an expression for matrix elements and expectation values. This is done analyzing the zeros and poles of $d\log[\sin(L\Omega_0)]$ and writing this function as an infinite  sum over its poles. The fact that $d\log(\sin(L\Omega_0)$ behaves as $d\log(\sin(L\lambda))$ at infinity, a differential for which such a series converges shows that this procedure is legitimate and produces the formula:
\begin{align}
&d\log(\sin(L\Omega_0(\lambda)/\hbar))=\oint \frac{1}{\lambda-\lambda'}  d\log[\sin(L\Omega_0(\lambda')/\hbar )]
\end{align}
The right hand side receives contributions from each point, $\sigma_i,$ at which $\sin(L\Omega_0(\sigma_i)/\hbar )=0$. We  exclude in the set $\sigma_i$ all the points $\lambda_i$ which are also solutions of $\sin(L\Omega_0/\hbar)=0,$ an assumption which was motivated in the previous paragraph.

\subsection{ Convergence of Quantum Matrix Elements to Classical Averages\label{SectionQExpToCAverage}}

Once the wavefunction has been written in separated coordinates, it is possible to compute the expectation values and matrix elements within this representation:
\begin{align}
\<\{\lambda_i\}|f(\{\mu_i\})|\{\lambda_i'\}\>=\int f(\{\mu_i\})\chi_{\{\lambda_i\}}(\{\mu_j\})\chi_{\{\lambda'_i\}}(\{\mu_j\})\prod_id\mu_i\prod_{i<j}(\mu_i-\mu_j).\label{obviousstuff}
\end{align}    
In the case where the matrix element computed is an expectation value, $\lambda'_i=\lambda_i$ this formula turns into the classical equation for averaged quantities\cite{Smirnov:Babelon:Bernard:Qauntization:Solitons,Smirnov:Babelon:Defoned:Hyperlliptic,Smirnov:Quasi-Classical:qKdV}, Eq. (\ref{classicalavging}), upon substitution of the quasiclassical expression for the wave function, which is a consequence  of  Eq (\ref{Qresult}), and making a saddle point approximation. 

To show the convergence of classical expectation values into classical averages, and to obtain more generally a semiclassical formula to compute matrix elements, let us first analyze the residue of the pole at each one of the zeros of $\sin(L\Omega_0/\hbar )$. At the points $\sigma_i$, the function $\Omega_0$  is regular and so $\sin(L\Omega_0/\hbar )$ has a simple zero and thus $d\log[\sin(L\Omega_0(z')/\hbar )]$ has a simple pole of residue $1.$ At the branch points, $\lambda_i$, the function $\Omega_0$ has the form $\sqrt{\lambda-\lambda_i}+\imath k\pi $, and so  $d\log[\sin(L\Omega_0(z')/\hbar )]$  has a simple pole of residue $1/2$.    We thus obtain:

\begin{align}
&d\log(\sin(L\Omega_0(\lambda)/\hbar ))=d\log\left[\sqrt{\prod_i(\lambda-\lambda_i)}\prod_j\left(1-\frac{\lambda-\lambda_1}{\sigma_j-\lambda_1}\right)\right]
\end{align}
which leads to
\begin{align}
\sin(L\Omega_0(\lambda)/\hbar )=\sqrt{(\lambda-\lambda_1)\prod_{i>1}\left(1-\frac{\lambda-\lambda_1}{\lambda_i-\lambda_1}\right)}\prod_j\left(1-\frac{\lambda-\lambda_1}{\sigma_j-\lambda_1}\right)\lim_{\tilde \lambda\to\lambda_1}\left( \frac{\Omega_0(\tilde \lambda)}{\sqrt{\tilde \lambda-\lambda_1}}\right).\label{RiemannSin}
\end{align}
 
Combining then Eqs. (\ref{xproperty}) and (\ref{Qresult}), $Q$\ being easily related to $\chi$  through Eq. (\ref{QchiRelation}), one obtains the following equation for the matrix element in Eq. (\ref{obviousstuff}) as an $N$ dimensional integral over a set of $N$\ variables $\bm{\mu}=\{\mu_i\}_{i=1}^N$:\begin{align}&\<\bm{\lambda}-\delta \boldsymbol{\lambda}|f(\bm \mu)|\bm{\lambda}+\delta \boldsymbol{\lambda}\>=\\&=\iint\dots\int  \frac{ e^{-\frac{ 2\imath }{ L\hbar c}\sum_j\int ^{\mu_j}\left( \Omega_0(\gamma,\bm{\lambda}+\delta \boldsymbol{\lambda})+\bar \Omega_0( \gamma,\bm{\lambda}+\delta \boldsymbol{\lambda})-2\pi L j\hbar \right)d\mu_j }}{\Gamma_{\boldsymbol{\lambda}+\delta \boldsymbol{\lambda}}( \bm\mu)\Gamma_{\boldsymbol{\lambda}-\delta \boldsymbol{\lambda}}(\sigma \bm\mu)  }   f(\bm \mu)\Delta(\bm{\mu})d\bm \mu, \end{align}
where
\begin{align}
\Gamma_{\bm \lambda}(\bm{\mu})=e^{\sum_j\int^{\mu_j}x(\gamma)d\gamma}, \quad \Delta(\bm{\mu})=\prod_{i<j}(\mu_i-\mu_j)\label{GammaDef}
\end{align}
The integral is now computed in the saddle point approximation. The saddle point equations dictate:
\begin{align}
 \Omega_0(\gamma,\bm{\lambda}+\delta \boldsymbol{\lambda})+\bar \Omega_0(\gamma,\bm{\lambda}+\delta \boldsymbol{\lambda})=2\pi L j\hbar.
\end{align} 
For $\delta \bm{\lambda}=0$ the solutions to these equations are the $\sigma_i$ and the the cuts (the segments $[\lambda_{2i-1},\lambda_{2i}])$, the former being a consequence of the definition of $\sigma_i$ and the latter being a consequence of the fact that $\Omega_0$ is imaginary on the cuts.
When $\delta \bm\lambda$ is non-zero, the location of the saddle points which were previously $\sigma_i$'s are perturbed and while the cuts are replaced by the segments \begin{align}[\lambda_{2i-1}+\delta \lambda_{2i-1},\lambda_{2i}+\delta\lambda_{2i}]\cap[\lambda_{2i-1}-\delta \lambda_{2i-1},\lambda_{2i}-\delta\lambda_{2i}] \end{align} One thus obtains the result as a $g+1$ dimensional integral over the $g+1$ segments above each one covered by the variable $\mu_i.$ We write for this set of variables $\bm{\mu}_{g+1}\equiv\{\mu_i\}_{i=1}^{g+1}$ and obtain, aided by Eq. (\ref{RiemannSin}):
\begin{align}
&\<\bm{\lambda}-\delta \boldsymbol{\lambda}|f(\bm \mu)|\bm{\lambda}+\delta \boldsymbol{\lambda}\>=\iint \dots\int \mathcal{G}({\boldsymbol{\lambda}+\delta \boldsymbol{\lambda}},\boldsymbol{\lambda}-\delta \boldsymbol{\lambda}) \frac{ f(\bm \mu_{g+1})}{\sqrt{\prod_{i,j}(\mu_i-\lambda_j)}}   d\bm \mu_{g+1}\Delta(\bm{\mu}_{g+1}),\label{QAveraging}
\end{align} 
where
\begin{align}
\mathcal{G}=\prod_{i=1}^\infty\frac{\sin\frac{ \Omega_0(\mu_i,\bm{\lambda}+\delta \boldsymbol{\lambda})+ \Omega_0(\mu_i,\bm{\lambda}-\delta \boldsymbol{\lambda})}{2L\hbar }}{\Gamma_{\boldsymbol{\lambda}+\delta \boldsymbol{\lambda}}(\mu_i)\Gamma_{\boldsymbol{\lambda}-\delta \boldsymbol{\lambda}}(\sigma \mu_i)}.
\end{align}
Note that for $\delta \bm\lambda=0$, the function $\mathcal{G}$ goes to $1$ due to Eq. (\ref{GammaDef}) and  Eq. (\ref{xproperty}). Normalizing the wave bra and ket states and setting $\delta \bm\lambda=0$ reproduces Eq.  (\ref{classicalavging}), showing that classical averages are equal to quantum expectation valuesas shown in Refrs. \cite{Smirnov:Babelon:Bernard:Qauntization:Solitons,Smirnov:Babelon:Defoned:Hyperlliptic,Smirnov:Quasi-Classical:qKdV}.  Indeed, this section comprised of a review of the  work presented originally in  the latter references.

\section{Quantum Ripples Over a Semiclassical Bosonic Shock }
We wish to find the momentum fluctuations of a Bose gas described the quantum nonlinear Schr\"{o}dinger equation out of equilibrium after the appearance of a shock. The shock is represented in Whitham theory or generalized hydrodynamics as the appearance of more moduli describing the Riemann surface in the former or the support of the Bethe roots in the latter. 
The density is described by $\<\hat\psi^\dagger\hat\psi\>,$ the classical analogue of which can be easily computed and then averaged in the oscillatory region to produce a semiclassical result for the density in the small $c$ limit. This procedure is, however unsatisfactory in the regions where there is a topological transition in the Riemann surface. Namely, the procedure fails in region where two $\lambda_i$'s meet and annihilate, consequently reducing the genus of the Riemann surface by $1$, or when a pair of $\lambda_i$'s appear from the same point and thus increasing the genus of the Riemann surface by $1$. In those regions our procedure will be a  point splitting procedure where we shall first compute $\<\hat\psi^\dagger(x-\frac{y}{2})\hat\psi(x+\frac{y}{2})\>$  and infer from that the density by taking $y\to0$ at the end. As we shall see this procedure allows to get more information about the fine structure of the density than the straightforward approach of setting $y$ to $0$ from the outset. 

The point splitting procedure employed here leads naturally to consider the Wigner function, by Fourier transforming in the separation between the split points:
\begin{align}
W(x,p)=\int\left\<\hat\psi^\dagger\left(x+\frac{y}{2}\right)\hat\psi\left (x-\frac{y}{2}\right)\right\>e^{\frac{-\imath yp}{\hbar}} dy.
\end{align}
The density is then obviously $\rho(x)=\<\hat\psi^\dagger(x)\hat\psi(x)\>=\int W(x,p)\frac{dp}{2\pi}.$ The semiclassical expression for the Wigner function, as described by the prescription provided in Refrs. \cite{Smirnov:Babelon:Bernard:Qauntization:Solitons,Smirnov:Babelon:Defoned:Hyperlliptic,Smirnov:Quasi-Classical:qKdV}
and reviewed in section \ref{SectionQExpToCAverage}, is then given by:
\begin{align}
&\left\<\hat\psi^\dagger\left(x-\frac{y}{2}\right)\hat\psi\left (x+\frac{y}{2}\right)\right\>=\oint\mathcal{G}(\bm    \lambda^{\rm out},\bm \lambda^{\rm in}) \frac{\bar\psi_{\bm \lambda^{\rm out}}(\mu_i) \psi_{\bm \lambda^{\rm in}}(\mu_i)\prod_{i<j}(\mu_i-\mu_j)}{{\prod_{i,j}(\mu_i-\lambda^{}_j)^{1/2}}{}},
\end{align} 
where $\bm \lambda^{\rm in/out}\equiv \{\lambda_i(x\pm y/2)\}_{i=1}^{2g+2}$       and $\lambda_i=\frac{\lambda^{\rm in}+\lambda^{\rm out}}{2}$.  Now  according to Eq. (\ref{psiAsMus}) we may write this as:
\begin{align}
W(x,p)=\int dy\beta^{\rm in}\beta^{\rm out}\oint \frac{ \mathcal{G}(\bm    \lambda^{\rm out},\bm \lambda^{\rm in})e^{\frac{\imath}{\hbar }\int_{x-y/2}^{x+y/2} \left[p_0-p-\sum_i\mu_i(x')\right]dx' }\prod_{i<j}(\mu_i-\mu_j)d\mu_idy}{\prod_{i,j}(\mu_i-\lambda^{}_j)^{1/2}}.\label{WignerInmus}
\end{align}

At this point we would like to remark on the significance of the point splitting procedure. We note that in order for the expression for $W(x,p)$ in Eq. (\ref{WignerInmus}) to be a valid semiclassical approximation then $y$ appearing in  exponent should be much larger than the period of oscillations, since singularities that arise in field theory when two operators are brought next to each other are not guaranteed to be captured by semiclassics in the form given in Eq.  (\ref{WignerInmus}). As a result, we may replace the integral $\int_{x-y/2}^{x+y/2}\sum_i \mu_i (x')dx'$ by $y\<\sum_i\mu_i\>$, and we obtain:
\begin{align} 
W(x,p)=\int\<\bm    \lambda^{\rm out}|\bm \lambda^{\rm in}\>\beta^{\rm in}\bar \beta^{\rm out}e^{\imath\int_{x-y/2}^{x+y/2} \left[p_0-p-\<\sum_i\mu_i\>\right]dx' } dy,
\end{align}     
where $\<\bm    \lambda^{\rm out}|\bm \lambda^{\rm in}\>$ may be computed making use of Eq. (\ref{QAveraging}). One may then incorporate all the $\bm \lambda$ dependant factors into one term, denoted by $P(x)$:
\begin{align}
P(x)\equiv p_0(x)-\<\sum_i\mu_i\>+2\Re \sum_i\frac{\partial\lambda_i}{\partial x} \frac{\partial}{\partial\lambda_i}\log\left(\beta(\bm \lambda)\<\bm    \lambda^{\rm }|\bm \lambda'^{\rm } \>\right)|_{\bm \lambda'=\bm\lambda},
\end{align}
and write:
\begin{align} 
W(x,p)=|\beta(x)|^2\int e^{\imath\int_{x-y/2}^{x+y/2} \left(P(x')-p\right)dx' } dy.\label{WignerDumpster}
\end{align}     

If $P(x)$ in Eq. (\ref{WignerDumpster}) is regular then one may approximate $\int_{x-y/2}^{x+y/2} P(x')=P(x)y$ and then  $W(x,p)\sim |\beta|^2\delta(P(x)-p)$, and the density is simply $\rho(x)=|\beta(x)|^2$. However if $P(x)$ is small then one must write down $\int_{x-y/2}^{x+y/2} P(x')=P(x)y+P''(x)\frac{y^3}{12} .$ This leads, by simple integration, to a Wigner function which has an Airy function form:
\begin{align}
W(x,p)=\frac{|\beta(x)|^2}{P''(x)^{1/3}}   {\rm Ai}\left(\frac{2^{2/3}(P(x)-p)}{P''(x)^{1/3}} \right),
\end{align}
 valid around the point $x_0$. The density, in contrast to the Wigner function, does not change its form and is still equal to $|\beta(x)|^2$, as can be ascertained by first doing the  integral in $p$ and then in $y$ to compute the density from Eq. (\ref{WignerDumpster}). 

We now turn to study the average momentum . It is convenient to subtract the value of  $P(x)$ from the average momentum. The distribution function associated with momentum after subtraction of $P(x)$ is denoted by  $G(x,p)$ and is defined as follows:
\begin{align}
G(x,p)\equiv(P(x)-p) W(x,p)=|\beta(x)|^2\frac{(P(x)-p)}{P''(x)^{1/3}}   {\rm Ai}\left(\frac{2^{2/3}(P(x)-p)}{P''(x)^{1/3}} \right),\label{AiryFuncWigner}
\end{align}
 where the last equality is understood to be valid around $x_0$. We examine the expression (\ref{AiryFuncWigner}) in  detail  to extract the behavior of the density near a point of topological transition, even though the expression in Eq. (\ref{AiryFuncWigner}) does not apply to this case a-priori, since $P(x)$\ obtains an infinite derivative at such points, as can be ascertained by computing this factor explicitly around points of topological transition. At such points $P(x)$\ contains terms of the form $(x-x_0)^{n/2}\log(x-x_0)^{m/2}$ where $m$ and $n$ are integers and $x_0$ is the point of topological transition. 

Even though $P(x)$ has infinite derivative near the points of topological transitions of the Riemann surface, thinking about the  function $P(x)$\ as a curve in $x-p$ space given by the set of points of the form $(x,P(x)),$ one recognizes that the ill behavior of the function $P(x)$\ can be removed by applying a rotation in $x-p$ space. This line of though is especially useful since the  expression for $G(x,p)$ in Eq. (\ref{AiryFuncWigner})  has  symmetry with respect to such rotations in $x-p$ space. Indeed, consider such a rotation:
\begin{align}
\left(\begin{array}{c}x\\p\end{array}\right)\to\left(\begin{array}{cc}c &s \\-s &c\end{array}\right)\left(\begin{array}{c}x\\p\end{array}\right),
\end{align}  
where $c=\cos(\theta) $ and $s=\sin(\theta)$, and $\theta$\ is the rotation angle. We have 
\begin{align}
P'\to \frac{-s+cP'}{c+sP'}, \quad P''\to \frac{P''}{(c+sP')^3},\label{PdoubleprimeTrans}
\end{align} 
 the latter expression appearing in the argument of (\ref{AiryFuncWigner}) and as a prefactor. Let us define three points in phase space:
\begin{align}
\vec{r}_A=\left(\begin{array}{c}x \\ p\end{array}\right), 
\quad \vec{r}_B=\left(\begin{array}{c}x_0 \\ P(x_0)\end{array}\right),
\quad \vec{r}_C=\left(\begin{array}{c}x \\ P(x)\end{array}\right),
\end{align} 
with which the Argument (up to constant factors) of the Airy function  in Eq. (\ref{AiryFuncWigner})\ can be written as:
\begin{align}
\frac{(\vec{r}_A-\vec{r}_B)\times(\vec{r}_C-\vec{r}_B)}{(x-x_0)P''^{1/3}}.
\end{align}
This expression can be seen to be invariant under rotations. Indeed, the expression in the numerator is invariant under rotations due to the fact that the vector product, being an area, is invariant, while the denominator is invariant if $x-x_0$ is small, as in this case one may use the transformation law for differentials $dx\to (c+sP')dx$, which when combined with the second transformation law in  Eq. (\ref{PdoubleprimeTrans}) shows that the denominator is invariant. Our conclusion is then that the object $|P''(x_0)|^{1/3}W(x,p)$ has a  rotationally invariant form around the point $x_0$ and has the same Airy function form in any rotated frame.

 We wish then to rotate the frame to new coordinate $\tilde{x}$ and $\tilde{p}$:
\begin{align}
\left(\begin{array}{c}\tilde x\\\tilde p\end{array}\right)=\left(\begin{array}{cc}c &s \\-s &c\end{array}\right)\left(\begin{array}{c}x\\p\end{array}\right),
\end{align}  
such that $P'(x_0)=\frac{s}{c}$, such that the curve $(\tilde{x},\tilde{P}(\tilde{x})),$ which  is the rotation of the curve $(x,P(x))$, has zero derivative at $x_0$. Namely,  $\tilde{P}'(\tilde{x}_0)=0$. The function $\tilde{P}$ is defined by the relation $\tilde{P}(cx+sP)=-sx+cP(x). $  

We then compute  $\delta\rho_p(x)\equiv\int G(x,p)dp$, by assuming the Airy function form, Eq. (\ref{AiryFuncWigner}), in the rotated coordinates:
\begin{align}
G(x,p)=\frac{|\beta(x)|^2}{|\tilde P''(\tilde x)|^{1/3}}  (P(\tilde{x}(x,p)-\tilde p(x,p)) {\rm Ai}\left(\frac{2^{2/3}(\tilde P(\tilde{x}(x,p))-\tilde p(x,p))}{\tilde  P''(\tilde x)^{1/3}} \right).
\end{align} 
We may represent then $\delta\rho_p$ directly through the Airy integral.
\begin{align}
&\delta\rho_p(x)=\frac{|\beta|^2}{\tilde P''^{1/3}}\int dp dy(\tilde P(\tilde x)-\tilde p)e^{\imath[\frac{\tilde P''}{24}    y^3+(\tilde P(\tilde x)-\tilde p)y]}
\end{align}
It is now more convenient to write the $\tilde P(\tilde{x})-p$ as $-\imath \partial_y -\frac{\tilde{P}''}{8}y^2$ acting on the exponent which upon disposing of the full derivative leads to:
\begin{align}
&\delta\rho_p(x)=-\frac{\tilde P''^{2/3}|\beta|^2}{8}\int dp dyy^2e^{\imath[\frac{\tilde P''}{6}(\frac{1}{4}  y^2+3(\tilde x-\tilde{x}_0)^2)+(\tilde p_0-\tilde p)]y},\end{align}
where we have also made the approximation $\tilde{P}(\tilde{x})=\tilde{p}_0+\frac{\tilde{P}''}{2} \tilde{x}^2$. We make the  substitution $p_\pm=\tilde{x}-\tilde{x}_0-\frac{c}{P'' s}\pm \frac{y}{2}$, which yields:
\begin{align}
&\delta\rho_p(x)=-\frac{\tilde P''^{2/3}|\beta|^2}{8}\int    dp_+dp_-(p_+-p_-)^2e^{\imath\frac{\tilde P''}{6}(p_+^3-p_-^3)+\imath(p_+-p_-)\left(\frac{x-x_0}{s}-\frac{c^2}{2P''s^2}\right)}\end{align}
The last integral separates is easily computable due to the fact that the exponent separates, and the result is: 

\begin{align}&\delta\rho_p(x)=\frac{|\beta|^2}{4}{(l{\rm Ai}^2(l)-{\rm Ai}'^2}\left(l\right)),\label{AiryFinal}
\end{align}    
where:
\begin{align}
\tilde P''^{1/3}l=\frac{{x-x_0}}{s}-\frac{c^2}{2\tilde P''s^2}
\end{align}
which, given that  $P'=\frac{s}{c}$ and $\tilde P''=\frac{P'' c^2}{(c+P's)}=\frac{P''}{(1+P'^2)^{3/2}},$ reads more explicitly as:
\begin{align}
l=\frac{1+P'^2}{P'P''^{1/3}}\left(x-x_0   -\frac{1+P'^2}{2P''P'}\right).
\end{align}
For example, if $P(x)=\sqrt{2}\alpha^{3/2}\sqrt{x-x_0}$ then 
\begin{align}
l=\alpha(x-x_0)-\frac{1}{2}\alpha^4
\end{align}
\section{Conclusion}

To conclude we have described in this paper how the semiclassical limit can be taken in the quantum integrable system known as the Lieb-Liniger model or the quantum non-linear Schr\"odinger equation. It was shown that the generalized hydrodynamics method \cite{Casto:Alvaredo:Doyon:Yoshimura:GHD} leads to the Whitham approach as fleshed out by Krichever \cite{83:Krichever:Averaging}. We have also discussed how this link allows to compute in the semiclassical limit quantities such as the Wigner function or average momentum fluctuations relying here on advances made by Babelon, Bernard and Smirnov\cite{Smirnov:Babelon:Bernard:Qauntization:Solitons,Smirnov:Quasi-Classical:qKdV}. The results show that  the $c\to0$ limit of the Lieb-Liniger model treated here and the $c\to\infty,$ equivalent to free-fermions, share similar features as can be seen by comparing the result here, Eq. (\ref{AiryFinal}), to that of Ref. \cite{Bettelheim:Glazman}. It is the hope that with a more general approach to expectation values  it will be possible to compute averaged quantities such as momentum fluctuations for the entire range of $c$'s, by the methods advanced, e.g., in \cite{Bettelheim:Kondo}. The results of the current paper may then be useful to validate such results.
   
\section{Acknowledgement}
I\ would like to thank J\'er\^ome Dubail
for suggesting this work and for the following discussions. I\ would like to acknowledge money from  ISF\ grant number 
 1466/15. \appendix
 \section{The Monodromy Matrix }
The two columns of the Baker-Akhiezer matrix are solutions of the linear problem, Eq. (\ref{LinearProblems}).The monodromy matrix is another matrix which depends on $x,$ $t$ and the spectral parameter, the columns of which satisfy the same equations, and as such the columns of the monodromy matrix can be written as linear combinations of the columns of the Baker-Akhiezer matrix. Even though he first column that is already given in (\ref{BA11explicit}, \ref{BA21explicit}),  we write it here along with the rest of the matrix elements  by making use of (\ref{BAInvolution}) and
(\ref{alphaExpression}): \begin{align}
&\Psi_{11}(z)=\frac{\theta(\bm{A}(z)+\imath \bm{P} x +\imath\bm{E} t-\bm{D})\theta(\bm{D})}{\theta(\boldsymbol{A}(z)-\bm{D})\theta(\imath \bm{P} x +\imath\bm{E} t-\bm{D})}e^{\frac{\imath}{\hbar } \left[x( \Omega_0(z)+p_0/2)-\frac{t}{m\hbar} (\Omega_1+mE_0/4)\right]}\label{secondBA11explicit}\\
&\Psi_{21}(z)=\frac{\theta(\bm{A}(z)+\imath \bm{P} x +\imath\bm{E} t+\bm{r}-\bm{D})\theta(\bm{D})}{\theta(\boldsymbol{A}(z)-\bm{D})\theta(\imath \bm{P} x +\imath\bm{E}t-\bm{D})}e^{\frac{\imath}{\hbar } \left[x( \Omega_0(z)-p_0/2)-\frac{t}{m\hbar} (\Omega_1-mE_0/4)\right]+\Omega_{-1}+\imath \arg(\theta(\bm{D}+\bm{r}))}\label{secondBA21explicit}\\&
\Psi_{12}(z)=\frac{\theta(-\bm{A}(z)+\imath \bm{P} x +\imath\bm{E} t-\bm{r}-\bm{D})\theta(\bm{D})}{\theta(\boldsymbol{A}(z)+\bm{r}+\bm{D})\theta(\imath \bm{P} x +\imath\bm{E} t-\bm{D})}e^{-\frac{\imath}{\hbar } \left[x( \Omega_0(z)-p_0/2)-\frac{t}{m\hbar} (\Omega_1-mE_0/4)\right]}\label{BA12explicit}\\
&\Psi_{22}(z)=\frac{\theta(-\bm{A}(z)+\imath \bm{P} x +\imath\bm{E} t-\bm{D})\theta(\bm{D})}{\theta(\boldsymbol{A}(z)+\bm{r}+\bm{D})\theta(\imath \bm{P} x +\imath\bm{E}t-\bm{D})}e^{-\frac{\imath}{\hbar } \left[x( \Omega_0(z)+p_0/2)-\frac{t}{m\hbar} (\Omega_1+mE_0/4)\right]-\Omega_{-1}+\imath \arg(\theta(\bm{D}+\bm{r}))}\label{BA22explicit}
\end{align}

The monodromy $\bm{T}(z,x)$ is defined as the matrix, the columns of which  solve (\ref{LinearProblems}), while having the initial conditions $\bm{T}(z,x)=\mathds{1}.$ The elements of the matrix are thus given by:
\begin{align}
\bm{T}= \Psi \left( \begin{array}{cc} 
\alpha & \gamma  \\
\beta & \delta 
 \end{array}\right)
\end{align}
where the matrix elements $\alpha,$ $\beta, $ $\gamma $ and $\delta$ are given at $t=0$ by:
\begin{align}
&\alpha=\frac{\theta(\bm{A}(z) +\bm{D})\theta(\boldsymbol{A}(z)-\bm{D})e^{-\Omega_{-1}}}{\theta(\bm{A}(z) +\bm{D})\theta(\boldsymbol{A}(z)-\bm{D})e^{-\Omega_{-1}}-\theta(\boldsymbol{A}(z)+\bm{r}+\bm{D})\theta(\bm{A}(z) +\bm{r}-\bm{D})e^{\Omega_{-1}}} \\
&\beta=-\frac{\theta(\bm{A}(z) +\bm{r}-\bm{D})\theta(\boldsymbol{A}(z)+\bm{r}+\bm{D})e^{\Omega_{-1}}}{\theta(\bm{A}(z) +\bm{D})\theta(\boldsymbol{A}(z)-\bm{D})e^{-\Omega_{-1}}-\theta(\boldsymbol{A}(z)+\bm{r}+\bm{D})\theta(\bm{A}(z) +\bm{r}-\bm{D})e^{\Omega_{-1}}}  \\
&\gamma=e^{-\imath \arg(\theta(\bm{D}+\bm{r}))}\frac{\theta(\boldsymbol{A}(z)+\bm{r}+\bm{D})\theta(\boldsymbol{A}(z)-\bm{D})}{\theta(\bm{A}(z)+\bm{r}-\bm{D})\theta(\boldsymbol{A}(z)+\bm{r}+\bm{D})e^{\Omega_{-1}}-\theta(\boldsymbol{A}(z)-\bm{D})\theta(\bm{A}(z)+\bm{D})e^{-\Omega_{-1}}} \\
&\delta=-\gamma         
\end{align} 
therefore  the off-diagonal elements of $\bm{T}(z,L),$ where $L$ is a period of the solution $\psi$, are given by ($\imath$  is missing somewhere since we have to have $T^*(\lambda)=\sigma_y T(\lambda)\sigma_y):$
\begin{align}
&T_{21}(z,L)=\frac{\theta(\boldsymbol{A}(z)+\bm{r}-\bm{D})\theta(\bm{A}(z) +\bm{D})e^{\imath(\arg(\theta(\bm{D}+\bm{r}))-Lp_0/2)}}{\theta(\bm{A}(z) +\bm{D})\theta(\boldsymbol{A}(z)-\bm{D})e^{-\Omega_{-1}}-\theta(\boldsymbol{A}(z)+\bm{r}+\bm{D})\theta(\bm{A}(z) +\bm{r}-\bm{D})e^{\Omega_{-1}}}  \times\\&\times\left[ e^{-\imath L \Omega_0}- e^{\imath L \Omega_0}\right]\\
&T_{12}(z,L)=\frac{\theta(\boldsymbol{A}(z)+\bm{r}+\bm{D})\theta(\boldsymbol{A}(z)-\bm{D})e^{-\imath(\arg(\theta(\bm{D}+\bm{r}))-Lp_0/2)}}{\theta(\bm{A}(z)+\bm{r}-\bm{D})\theta(\boldsymbol{A}(z)+\bm{r}+\bm{D})e^{\Omega_{-1}}-\theta(\boldsymbol{A}(z)-\bm{D})\theta(\bm{A}(z)+\bm{D})e^{-\Omega_{-1}}}  \times\\&\times\left[ e^{-\imath L \Omega_0}- e^{\imath L \Omega_0}\right]
\end{align}
The matrix element $T_{12}$ us known as $\mathcal{B}$ and its zeros are the one set of variables in the separation of values method. The zeros are here seen to be the zeros of $\theta(\bm{A}-\bm{D})$ and of $\theta(\bm{A}+\bm{D}+\bm{r})$ as well as the zeros of $ e^{-\imath L \Omega_0}- e^{\imath L \Omega_0}.$
 The zeros of $\theta(\bm{A}-\bm{D})$ lie on the real axis of the Riemann surface and the zeros of $\theta(\bm{A}+\bm{D}+\bm{r})$ are simply their reflection under $\sigma. $ The zeros of $ e^{\imath L \Omega_0}- e^{-\imath L \Omega_0}=2\imath\sin(L\Omega_0)$ lie typically on the real axis. For large $\lambda$ this function has the form $2\imath\sin(Lz)$, namely the zeros are sitting at $\frac{\pi n}{L}. $

We consider also the trace of the monodromy matrix. It is given by:
\begin{align}
&\tr(\bm{T})=2\cos\left(  L \Omega_0(z)\right),
\end{align}
under the assumption that $Lp_0/2\in 2\pi\mathds{Z}$ (which is required in order for the wave function $\psi$ to be periodic). Eq. (\ref{trTisCos1}) is derived similarly.

\bibliographystyle{unsrt}

\bibliography{mybib}

\begin{thebibliography}{10}

\bibitem{Das:Solitons}
A.~Das.
\newblock {\em Integrable Models}.
\newblock World Scientific, Singapore, 1989.

\bibitem{Faddeev:Book:Hamiltonian:Methods}
Ludwig Faddeev and Leon Takhtajan.
\newblock {\em Hamiltonian methods in the theory of solitons}.
\newblock Springer Science \& Business Media, 2007.

\bibitem{100:Bettelheim:Agam:Wieg}
O.~Agam, E.~Bettelheim, P.~W. Wiegmann, and A.~Zabrodin.
\newblock Viscous fingering and the shape of an electronic droplet in the
  quantum hall regime.
\newblock {\em Phys. Rev. L.}, 88 (23):236801(4), 2002.

\bibitem{Soliton:Book}
S.~Novikov, S.~V. Manakov, L.~P. Pitaevski{\v{\i}}, and V.E. Zakharov.
\newblock {\em Theory of {S}olitons: The Inverse Scattering Method}.
\newblock Consultants Bureau, New-{Y}ork and {L}ondon, 1984.

\bibitem{Whitham:Book}
G.~B. Whitham.
\newblock {\em Linear and Nonlinear Waves}.
\newblock Wiley, New York, 1974.

\bibitem{89:Kodama:Bloch}
A.~M. Bloch and Y.~Kodama.
\newblock Dispersive regularization of the whitham equations for the toda
  lattice.
\newblock {\em SIAM J. of App. Math}, 52 (4):909--928, 1992.

\bibitem{Carroll:Remarks:On:Whitham}
R.~{Carroll}.
\newblock {Remarks on the Whitham equations}.
\newblock {\em arXiv:solv-int/9511009}, 1995.

\bibitem{88:Flaschka:KdV:Avging}
H.~Flaschka, M.~G. Forest, and D.~W. McLaughlin.
\newblock Multiphase averaging and the inverse spectral solution of {KdV}.
\newblock {\em Comm. Pure. Appl. Math.}, 33:739--784, 1980.

\bibitem{Gurevich:Pitavsk}
A.~V. Gurevich and L.~P. Pitaevski{\v{\i}}.
\newblock Nonstationary structure of a collisionless shock wave.
\newblock {\em Sov. Phys. JETP}, 38 (2):291--297, 1974.

\bibitem{83:Krichever:Averaging}
I.~M. Krichever.
\newblock Method of averaging two-dimensional integrable equations.
\newblock {\em Functional Analysis and its Applications}, 22(3):200--213, 1988.

\bibitem{Casto:Alvaredo:Doyon:Yoshimura:GHD}
Olalla~A. {Castro-Alvaredo}, Benjamin {Doyon}, and Takato {Yoshimura}.
\newblock {Emergent Hydrodynamics in Integrable Quantum Systems Out of
  Equilibrium}.
\newblock {\em Physical Review X}, 6(4):041065, Oct 2016.

\bibitem{Bettelheim:Glazman}
E.~{Bettelheim} and L.~{Glazman}.
\newblock {Quantum Ripples Over a Semiclassical Shock}.
\newblock {\em Physical Review Letters}, 109(26):260602, 2012.

\bibitem{belokolos:Bobenko:Algebro:Geometrical:Integrable}
Eugene~D Belokolos.
\newblock {\em Algebro-geometric approach to nonlinear integrable equations}.
\newblock Springer, 1994.

\bibitem{132:Krich:Intgr:AlgGeo}
I.~M. Krichever.
\newblock Integration of nonlinear equations by the methods of algebraic
  geometry.
\newblock {\em Func. Anal. Appl.}, 11:12--26, 1977.

\bibitem{84:Dubrovin:Algebr:Geome}
B.~A. Dubrovin.
\newblock Theta functions and non-linear equations.
\newblock {\em Russian Math. Surveys}, 36:2:11--92, 1981.

\bibitem{Korepin:Bogoliubov:Izergin:Quantum:Inverse:Scattergin}
Vladimir~E Korepin, Nicholay~M Bogoliubov, and Anatoli~G Izergin.
\newblock {\em Quantum inverse scattering method and correlation functions},
  volume~3.
\newblock Cambridge university press, 1997.

\bibitem{Smirnov:Babelon:Defoned:Hyperlliptic}
O.~{Babelon}, D.~{Bernard}, and F.~A. {Smirnov}.
\newblock {Form factors, KdV and Deformed Hyperelliptic Curves}.
\newblock {\em Nuclear Physics B Proceedings Supplements}, 58:21--33, 1997.

\bibitem{Sklyanin:SoV}
E.~K. Sklyanin.
\newblock {\em J. Soviet Math.}, 31:3417--3431, 1985.

\bibitem{Smirnov:Babelon:Bernard:Null:Vectors}
O.~{Babelon}, D.~{Bernard}, and F.~A. {Smirnov}.
\newblock {Null-Vectors in Integrable Field Theory}.
\newblock {\em Communications in Mathematical Physics}, 186:601--648, 1997.

\bibitem{Smirnov:Babelon:Bernard:Qauntization:Solitons}
O.~{Babelon}, D.~{Bernard}, and F.~A. {Smirnov}.
\newblock {Quantization of Solitons and the Restricted Sine-Gordon Model}.
\newblock {\em Communications in Mathematical Physics}, 182:319--354, 1996.

\bibitem{Smirnov:Quasi-Classical:qKdV}
F.~A. {Smirnov}.
\newblock {Quasi-classical Study of Form Factors in Finite Volume}.
\newblock {\em arXiv:hep-th/9802132}, 1998.

\bibitem{Bettelheim:Kondo}
E.~{Bettelheim}.
\newblock {Towards a non-equilibrium Bethe ansatz for the Kondo model}.
\newblock {\em Journal of Physics A Mathematical General}, 48(16):165003, 2015.

\end{thebibliography}

\end{document}